\documentclass[a4paper,twocolumn,english,aps,prl,reprint,superscriptaddress,showpacs,showkeys,longbibliography]{revtex4-1}
\usepackage[latin9]{inputenc}
\usepackage{babel}
\usepackage{mathtools}
\usepackage{amsmath}
\usepackage{amssymb}
\usepackage{graphicx}
\usepackage{float}
\usepackage[colorlinks=true, linkcolor=red, citecolor=blue, urlcolor=blue]{hyperref}
\usepackage[usenames,dvipsnames]{color}
\usepackage{bm}
\usepackage{bbold}
\usepackage{braket}
\usepackage{comment}
\usepackage{dcolumn,extarrows}
\usepackage{soul}
\usepackage{footmisc}
\usepackage{pgf,tikz}

\definecolor{mygreen}{rgb}{0,0.5,0}
\definecolor{myblue}{rgb}{0,0,0.75}
\definecolor{mymagenta}{cmyk}{0,1,0,0.12}

\newcommand{\minus}{
  \setbox0=\hbox{-}
  \vcenter{
    \hrule width\wd0 height \the\fontdimen8\textfont3
  }%
}

\def\inner(#1,#2,#3,#4,#5,#6){\ensuremath\left(\begin{array}{ccc} #1 & #2 & #3 \\ #4 & #5 & #6 \end{array}\right)}

\def\innerv(#1,#2,#3,#4,#5,#6){\ensuremath\left\{\begin{array}{ccc} #1 & #2 & #3 \\ #4 & #5 & #6 \end{array}\right\}}

\definecolor{mygreen}{rgb}{0,0.5,0}\definecolor{myblue}{rgb}{0,0,0.75}\definecolor{mymagenta}{cmyk}{0,1,0,0.12}

\newcommand{\bla}[1]{\left( #1 \right)}
\newcommand{\blb}[1]{\left[ #1 \right]}

\makeatother

\begin{document}

\title{Protecting quantum spin coherence of nanodiamonds in living cells}
\author{Q.-Y.~Cao}
\author{P.-C.~Yang}
\email{pengchengyang@hust.edu.cn}
\author{M.-S.~Gong}
\author{M.~Yu}
\affiliation{School of Physics and Wuhan National Laboratory for Optoelectronics, Huazhong University of Science and Technology, Wuhan 430074, China}
\affiliation{International Joint Laboratory on Quantum Sensing and Quantum Metrology, Huazhong University of Science and Technology, Wuhan 430074, China}
\author{A.~Retzker}
\affiliation{Racah Institute of Physics, The Hebrew University of Jerusalem, Jerusalem, 91904 Givat Ram, Israel}
\author{M.~B.~Plenio}
\affiliation{Institut f\"{u}r Theoretische Physik $\&$ IQST, Albert-Einstein Allee 11, Universit\"{a}t Ulm, D-89081 Ulm, Germany}
\affiliation{International Joint Laboratory on Quantum Sensing and Quantum Metrology, Huazhong University of Science and Technology, Wuhan 430074, China}
\author{C.~M\"{u}ller}
\author{N.~Tomek}
\author{B.~Naydenov}
\author{L.~P. McGuinness}
\affiliation{Institut f\"{u}r Quantenoptik $\&$ IQST, Albert-Einstein Allee 11, Universit\"{a}t Ulm, D-89081 Ulm, Germany}
\author{F.~Jelezko}
\affiliation{Institut f\"{u}r Quantenoptik $\&$ IQST, Albert-Einstein Allee 11, Universit\"{a}t Ulm, D-89081 Ulm, Germany}
\affiliation{International Joint Laboratory on Quantum Sensing and Quantum Metrology, Huazhong University of Science and Technology, Wuhan 430074, China}
\author{J.-M.~Cai}
\email{jianmingcai@hust.edu.cn}
\affiliation{School of Physics and Wuhan National Laboratory for Optoelectronics, Huazhong University of Science and Technology, Wuhan 430074, China}
\affiliation{International Joint Laboratory on Quantum Sensing and Quantum Metrology, Huazhong University of Science and Technology, Wuhan 430074, China}

\begin{abstract}
Due to its superior coherent and optical properties at room temperature, the nitrogen-vacancy (N-$V$) center in diamond has become a promising quantum probe for nanoscale quantum sensing. However, the application of N-$V$ containing nanodiamonds to quantum sensing suffers from their relatively poor spin coherence times.
Here we demonstrate energy efficient protection of N-$V$ spin coherence in nanodiamonds using concatenated continuous dynamical decoupling, which exhibits excellent performance with less stringent microwave power requirement. When applied to nanodiamonds in living cells we are able to extend the spin coherence time by an order of magnitude to the $T_1$-limit of up to $30\mu$s. Further analysis demonstrates concomitant  improvements of sensing performance which shows that our results provide an important step towards in vivo quantum sensing using N-$V$ centers in nanodiamond.
\end{abstract}
\date{\today}

\pacs{42.50.Dv, 07.55.Ge, 03.67.-a, 42.50.-p}

\maketitle

\section{Introduction}

Nitrogen-Vacancy (N-$V$) centers in diamond exhibit stable fluorescence and have a spin triplet ground state, which can be coherently manipulated by microwave fields \cite{Doherty13}. Observation of spin-dependent fluorescence provides an efficient way to readout the spin state of N-$V$s. The energy splitting of the N-$V$ spin depends on physical parameters, such as magnetic field \cite{Maz_08_NL,Bal_08_NL,Tay_08_NP}, electric field \cite{Dol_11_NP,Dol_14_PRL}, temperature \cite{Acosta_2010_PRL, Kucsko_Nature_2013,Toy_13_PNASU,Neu_13_NL} and pressure \cite{Doherty_14_PRL,Cai NC14}. A variety of quantum sensing protocols for precise measurement of these physical parameters in different scenarios have been developed \cite{Schir14,Degen17,Hirose,Fang2013,Joas2017,Stark2017,Hall16,Cooper2014,Schmitt2017,Boss2017,Liu2017,Hsieh2019,Yip2019,Lesik2019}. These protocols are all based on determining the N-$V$ spin energy splitting, which is why the measurement sensitivity is limited by the N-$V$ spin coherence time.

Spin coherence in bulk diamond is mainly affected by surrounding electronic impurities (P1 centers) and nuclear spins (natural abundance of $^{13}$C isotope). The spin reservoir can be eliminated by using isotopically engineered high-purity type IIa diamond \cite{Bala09_NM}. In order to mitigate the influence of any residual impurities, pulsed dynamical decoupling has been widely exploited to prolong spin coherence time \cite{Lange_Sci_2010,Nay_PRB_2011,Knowl14}. Its excellent performance when applied to N-$V$s in bulk diamond is a result of the quasi-static characteristics of the spin reservoir in bulk diamond and the high available microwave power. Unfortunately, these two factors may not be satisfied for N-$V$ centers in nanodiamonds, which are required for sensing applications \emph{in vivo}. N-$V$s contained within nanodiamonds typically exhibit poor spin coherence time, which has been attributed to nanodiamond surface spin noise and electric charge noise that include prominent high frequency components. Preserving the coherence of N-$V$s in nanodiamonds becomes even more problematic when they are located in biological environments which presents additional noise sources. The microwave power available to decouple N-$V$s within living cells can be limited by the large distance between microwave antenna and nanodiamond, and the damaging effects that microwave absorption may have on biological tissue. Therefore, the development of an energy efficient strategy to prolong coherence time of N-$V$s in nanodiamond under the constraint of limited microwave power represents a significant challenge for efficient quantum sensing protocols for biology and nanomedicine \cite{Bala14,Mcgui11,Kucsko_Nature_2013,Sage_Nature_2013,Wu2016,Kurizki}.

In this work, we address this key challenge with the implementation of concatenated continuous dynamical decoupling (CCDD), which employs a microwave driving field consisting of suitably engineered multi-frequency components \cite{Cai_NJP_2012,Cohen16,Farf17}, to prolong coherence time of N-$V$s contained within nanodiamonds. The purpose of the main frequency component of the microwave
drive is to suppress fast environmental noise, while the other weaker frequency components compensate power fluctuations in the main frequency component. The key advantage of CCDD compared to pulsed schemes is that the decoupling efficiency achievable at the same average power (as quantified by the effective Rabi frequency $\bar{\Omega}^2=\langle \Omega^2 \rangle$ averaged over time) is predicted theoretically to be superior to pulsed schemes \cite{Kurizki}. We demonstrate experimentally that CCDD achieves a performance that significantly exceeds that of pulse dynamical decoupling strategies given the same microwave energy consumption. We show that CCDD prolongs coherence time of N-$V$s in nanodiamond up to tens of microseconds at which point it reaches the limit imposed by the N-$V$ spin relaxation time $T_1$ in these nanodiamonds inside living cells. Our result represents an important step towards the development of quantum sensing for {\it in vivo} applications with high achievable sensitivity, and also demonstrates wide applications of quantum control using amplitude and phase modulated driving field \cite{Shu2018,Yu2019}.

\section{Protecting coherence of N-$V$ spin in nanodiamond}


%
\begin{figure}[b]
\centering
\begin{minipage}{9cm}
\hspace{-0.6cm}
\includegraphics[width=8.8cm]{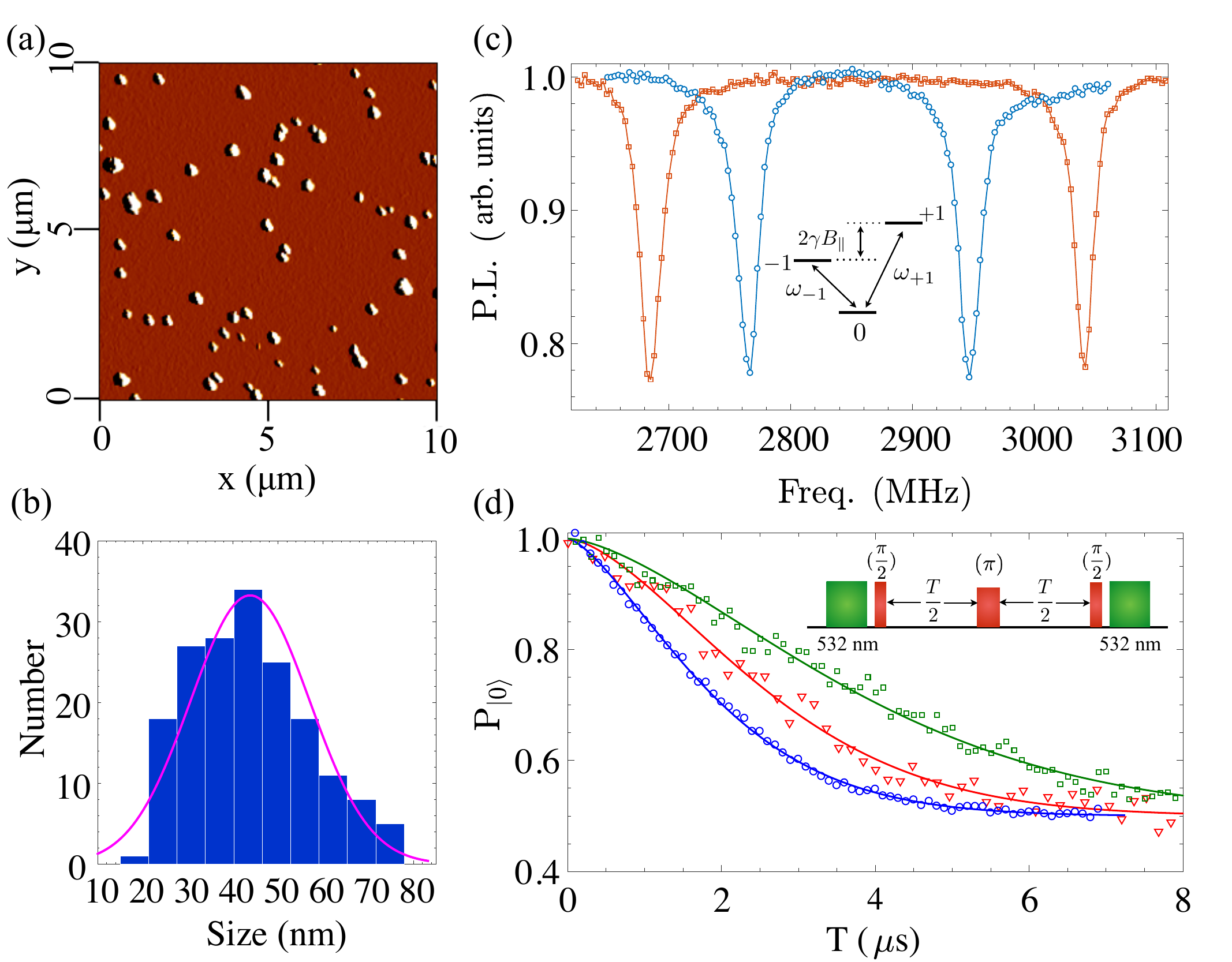}
\end{minipage}
\caption{Characteristics of N-$V$s in nanodiamond. \textbf{(a)} Atomic force microscope image of nanodiamonds deposited on a mica plate. \textbf{(b)} Histogram of nanodiamond sizes that are predominantly within 25-61nm ($43\pm18$ nm). \textbf{(c)} Typical ODMR measurement of a nanodiamond N-$V$ center with an applied magnetic field $B_{\parallel}=$ 37.8G (\protect\scalebox{0.6}{\color{Red}$\square$}) and 66G ({\color{Cyan}$\circ$}). \textbf{(d)} The time evolution of the spin state population $\mbox{P}_{\ket{0}}$ in spin echo experiment for three typical N-$V$s. By fitting the data with a function of $(1/2)[1+\exp(-(T/T_{SE})^{\alpha})]$, we estimate the parameters $[T_{SE},\alpha]$ as follows: $[ 2.142\pm 0.018\mu s, 1.448 \pm 0.026]$ (N-$V$1, {\color{Orange}$\circ$}), $[ 4.292\pm 0.133\mu s, 1.47\pm 0.10]$ (N-$V$2, \protect\scalebox{0.6}{\color{Cyan}$\square$}), $[ 2.990 \pm 0.083\mu s, 1.576 \pm 0.101]$ (N-$V$3, \protect\scalebox{0.8}{\color{Red}$\triangledown$}). }
\label{fig:basic}
\end{figure}
%

\begin{figure}[t]
\centering
\begin{minipage}{9cm}
\hspace{-0.6cm}
\includegraphics[width=9cm]{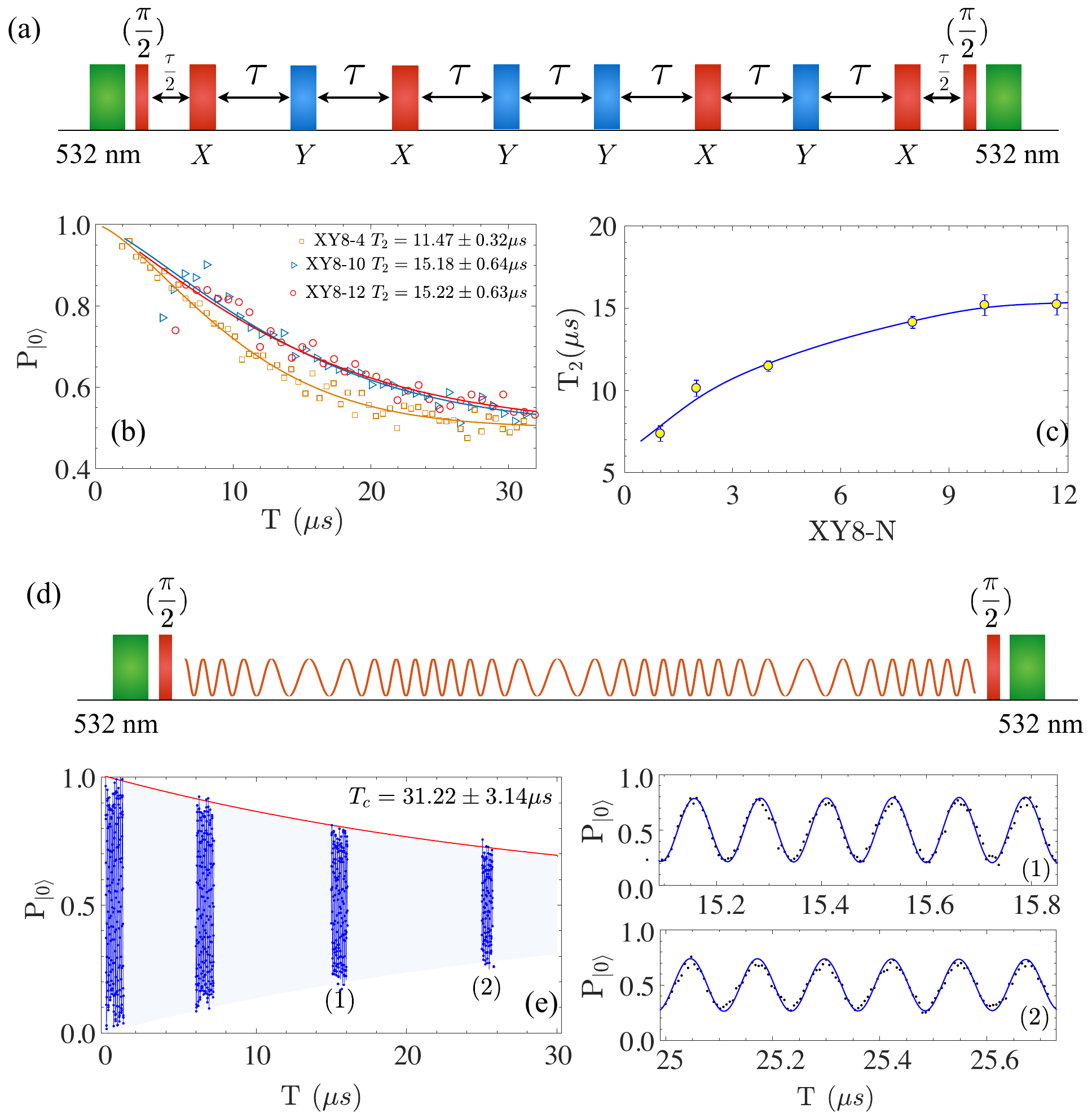}
\end{minipage}
\caption{The performance of concatenated continuous dynamical decoupling as compared with XY8 pulse dynamical decoupling. \textbf{(a)} XY8 pulse sequence. \textbf{(b)} Coherence time of nanodiamond N-$V$ is extended by XY8-N pulse sequence with the number of pulses up to 96 ($8\times N$). The Rabi frequency of $\pi$-pulse is $8.5$ MHz. The data is fitted with the decay function $(1/2)[1+\exp(-(T/T_2)^{\alpha})]$ which indicates the achievable coherence time of $T_2=15.22 \pm 0.63\mu s $. \textbf{(c)} shows the scaling of $T_2$ with the number of XY8 cycles $N$, which fits $(N^{-\beta}+T_{SE}/T_{1})^{-1}$ \cite{Medford_PRL_2012} with $\beta= 0.319\pm 0.064$. \textbf{(d)} Concatenated continuous dynamical decoupling scheme with phase modulated driving. {\bf(e)} The extended spin coherence time by CCDD reaches $T_C=31.22 \pm 3.14 \mu s$. The signal envelope is fitted by $(1/2)[1+\exp(-(T/T_C))]$. The driving parameters are $\Omega_1=8.06$ MHz, and $(\Omega_2/\Omega_1)=0.1$. The subpanels (1)-(2) show zoomed views over different measurement intervals. The applied magnetic field is $B_{\parallel}=508$ Gauss.}
\label{fig:XY8_CCD}
\end{figure}

In our experiment, we use nanodiamonds obtained by milling of HPHT diamond from Microdiamant with diameters of approximately $43\pm 18$ nm, as measured with atomic force microscope, see Fig.\ref{fig:basic}(a-b). We apply a static external magnetic field of strength $B$ along the N-$V$ axis, which leads to two allowed N-$V$ spin transitions ($m_s=0 \leftrightarrow m_s= +1$ and $m_s=0 \leftrightarrow m_s=-1$). The corresponding optically detected magnetic resonance (ODMR) measurement is shown in Fig.\ref{fig:basic}(c). We first characterize the coherence properties of single N-$V$ centers in nanodiamonds by performing spin echo measurements. Microwave control pulses are generated with an arbitrary waveform generator (AWG) which are amplified by a microwave amplifier. The power of microwave radiation determines the frequency of Rabi oscillation. Spin echo measurement were performed for several N-$V$s to determine the spin coherence time, three representative examples are shown in Fig.\ref{fig:basic}(d). The data is fitted by a decay function in the form of $\exp[-(t/T_{SE})^{\alpha}]$, where $T_{SE}$ denotes the spin echo coherence time. We extract the value of $\alpha$, the statistic of which  shows $\alpha\in[1.08,1.74]$, indicating decoherence is due to both slow and fast environmental fluctuations, see Appendix. Universal dynamical decoupling with a train of pulses, such as Carr-Purcell-Meiboom-Gill and XY8 sequences, may prolong spin coherence time by suppressing noise of low frequency \cite{Nay_PRB_2011,Knowl14}. Our Ramsey measurement under different magnetic field strengths shows that $T_2^*$ becomes longer as the magnetic field increases (see Appendix) which suggests that the noise in the present scenario is dominated by surface electric noise \cite{Jac16} rather than a slow spin bath.
We apply XY8-N pulse sequences to several N-$V$ centres using a train of $8N$ $\pi$-pulses, as shown in Fig.\ref{fig:XY8_CCD}(a). Fig.\ref{fig:XY8_CCD}(b) shows the measurement results obtained by applying up to $96$ $\pi$-pulses. The extended coherence time $T_2$ under dynamical decoupling increases as the number of XY8 cycles grows following the scaling $(N^{-\beta}+T_{SE}/T_{1})^{-1}$ with $N$ up to 12 \cite{Medford_PRL_2012}, see Fig.\ref{fig:XY8_CCD}(c). We also observe that the coherence time saturates and any further increase of number of  pulses does not necessary leads to a longer coherence time. Because the XY8 pulse sequence exhibits excellent pulse error tolerance \cite{Ahmed_2013_PRA}, the experiment observation suggests that the limited coherence time is likely due to fast noise dynamics and limited microwave power (pulse repetition rate).

%
%

%
To achieve high efficiency dynamical decoupling under the constraint of microwave power, we apply CCDD to protect spin coherence of nanodiamond N-$V$s. We begin by illustrating the basic idea of CCDD as applied to the $m_s=0 \leftrightarrow m_s=-1$ transition of a single N-$V$ spin \cite{Cai_NJP_2012,Cohen16,Farf17}. We would like to remark that the present scheme is applicable to many other two-level quantum systems. We introduce a microwave driving field with phase modulation \cite{Cohen16} as
\begin{equation}
\tilde{H}=(\Omega_{1}+\delta_x)\cos\blb{\omega_{0}t+2\bla{\frac{\Omega_{2}}{\Omega_{1}}}\sin(\Omega_{1}t)}\sigma_{x},\label{eq:ccdd_field}
\end{equation}
where $\sigma_{x}$ is Pauli operator, $\omega_{0}$ is the energy gap between $m_s=0$ and $m_s=-1$, $\Omega_{1}$ is the Rabi frequency as determined by microwave power, $\Omega_2$ denotes the ratio of phase modulation, and the fluctuation in microwave power is denoted as $\delta_x$. The effect of magnetic noise is suppressed by the driving field as long as the noise power density is small at frequency $\Omega_1$. In the interaction picture, the effective Hamiltonian can be written as \cite{Cohen16} $H_{I2}=-\bla{\Omega_{2}/2}\sigma_{z}+\delta_{x}\sigma_{x}$. As the phase control in the AWG is very stable, the fluctuation in $\Omega_2$ is negligible. As long as the power spectrum of $\delta_x$ is negligible at frequencies larger than $\Omega_2$, the noise will only lead to a second order effect, i.e., $ \bla{\delta_x^2/\Omega_2} \sigma_z$ \cite{Nati}. As compared with pulsed dynamical decoupling strategy, CCDD can achieve better performance with the same average microwave power \cite{Kurizki} as we will confirm in our experiment on nanodiamonds in living cells. In our experiment, a schematic of the microwave and readout sequence to implement CCDD is shown in Fig.\ref{fig:XY8_CCD}(d). A $\frac{\pi}{2}$-pulse prepares N-$V$ spin in a superposition state of $\ket{0}$ and $\ket{-1}$. The phase varying driving field as in Eq.(\ref{eq:ccdd_field}) is generated with an AWG and acts on the N-$V$ spin for time $T$. A final $\frac{\pi}{2}$-pulse maps spin coherence information into the state $\ket{0}$ population as measured by APD (avalanche photodiode gate). Fig.\ref{fig:XY8_CCD}(e) shows coherent oscillation by applying the CCDD scheme which leads to an extended coherence time $T_c=31.22\pm3.14 \mu s$, which is of the same order of the relaxation time $T_1$ (which we measured to be $87.35 \pm 7.50 \mu s$) as $T_2$ is limited by $T_1/2$ \cite{nir}.

%
\begin{figure}[b]
\centering
\begin{minipage}{9cm}
\hspace{-0.3cm}
\includegraphics[width=9cm]{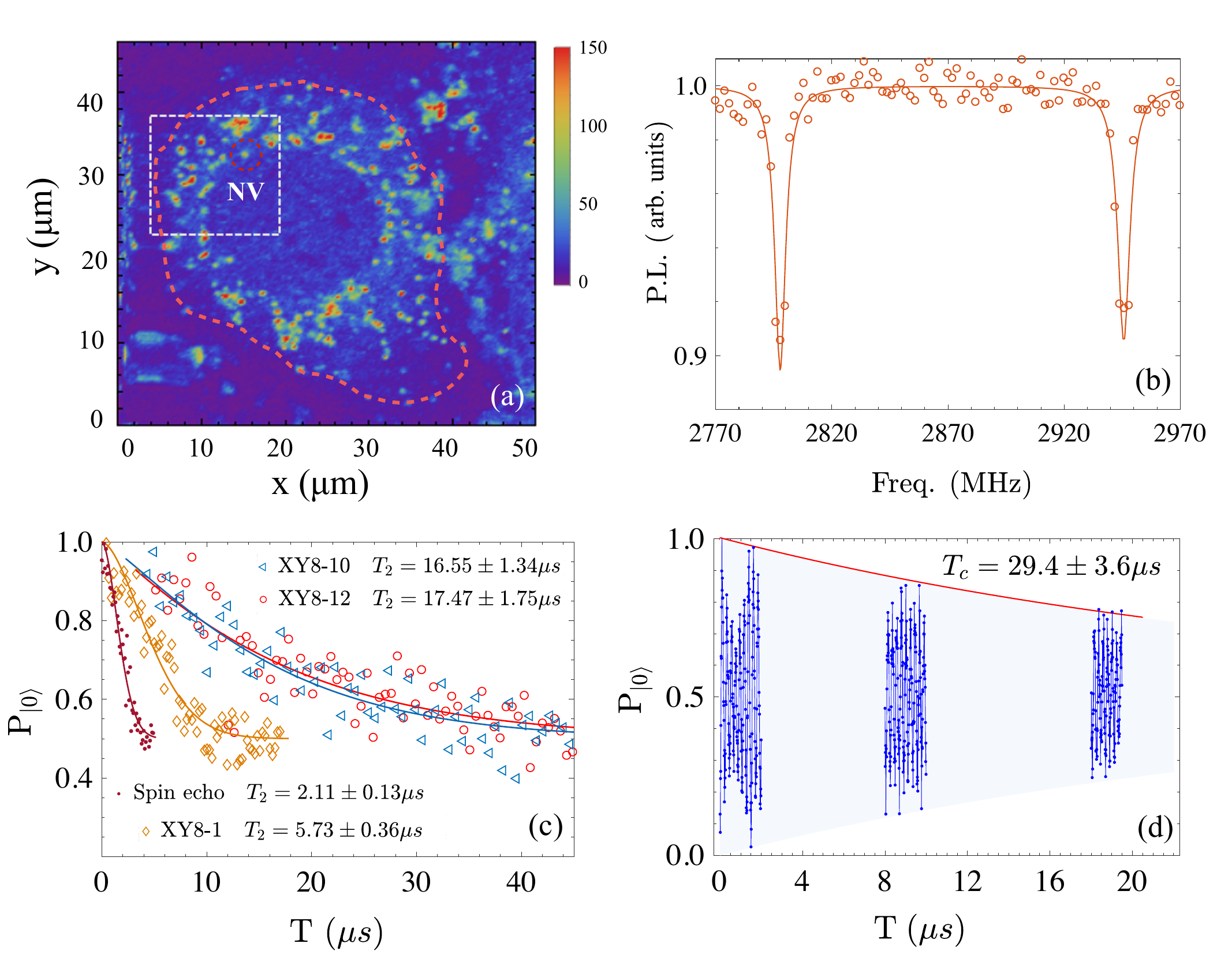}
\end{minipage}
\caption{Protecting quantum spin coherence in living cells. {\bf (a)} Confocal image of cell with an uptake of fluorescent nanodiamonds. The red curve indicates the cell boundary. The nanodiamonds in the dashed rectangular are located in the cytoplasm. The photon counts are in the unit of kcs/s. {\bf (b)} ODMR measurement of a nanodiamond N-$V$ center in cell with an applied magnetic field $B_{\parallel}=25$ Gauss. {\bf (c)} N-$V$ coherence time in living cell is extended by spin echo and XY8 pulse sequences (XY8-N) with the number of pulses up to 96. The Rabi frequency of $\pi$-pulse is $9.6$ MHz. {\bf (d).} The CCDD signal indicates a spin coherence time of $29.4\pm 3.6 \mu s$ with a Rabi frequency $\Omega_1=4.6$ MHz, and $\Omega_2=\Omega_1/10$. }
\label{fig:cellscan}
\end{figure}
%

The requirement of low microwave power is of practical importance for quantum sensing applications \emph{in vivo}, because microwave radiation is absorbed by biological tissues which may lead to heating and subsequent damage or denaturing of protein molecules. To demonstrate the performance of CCDD aimed at biosensing, we exploit the scheme to protect the spin coherence of N-$V$s in nanodiamonds up-taken by living cells. The cells we use in experiment are the NIH/3T3 cells that are adherent to the upper surface of the cover glass. To avoid the strong fluorescence of the nutrient solution, we replace it with phosphate buffered saline (PBS) which has almost no fluorescence and wash the cells 3 times to remove nanodiamonds not internalised by the cells. Fig.\ref{fig:cellscan}(a) shows a confocal scan image of the cell sample where fluorescence is gated around the N-$V$ emission spectrum. With membrane labelling and depth scan tomography, we clearly identify nanodiamonds that are taken up by the cells and located in the cell cytosol. We perform ODMR (Fig.\ref{fig:cellscan}b) and spin echo measurements (Fig.\ref{fig:cellscan}c) to characterize the properties of those nanodiamond N-$V$s in cells. We compare the performance of CCDD with XY8 pulsed dynamical decoupling for the protection of N-$V$ spin coherence in cells. Fig.\ref{fig:cellscan}(c-d) show the measurement data obtained by applying a microwave of the power that corresponds to a Rabi frequency of $9.6$ MHz and $4.6$ MHz in XY8 pulse dynamical decoupling and CCDD scheme respectively, the N-$V$ spin coherence time is extended up to $29.4\pm3.6\mu s$ by CCDD scheme while XY8 pulsed scheme reaches only $17.49\pm 1.43 \mu s$. We note that the advantage of CCDD scheme to become more prominent as the available microwave power is reduced and $T_1$ time is increased by nanodiamond material design. To characterise the damage effect of microwave radiation, we apply the experiment sequences and monitor the temperature increase of sample. The results suggest that the temperature increase due to the pulse sequences would be about 10~$^{\circ}$C more than that due to CCDD when achieving similar coherence times, see Appendix. Such a difference in temperature increase is expected to have a significant effect on biological tissue \cite{Woo2000,Shamis2012}.

%
%
\begin{figure}[t]
\centering
\begin{minipage}{9cm}
\hspace{-0.6cm}
\includegraphics[width=9cm]{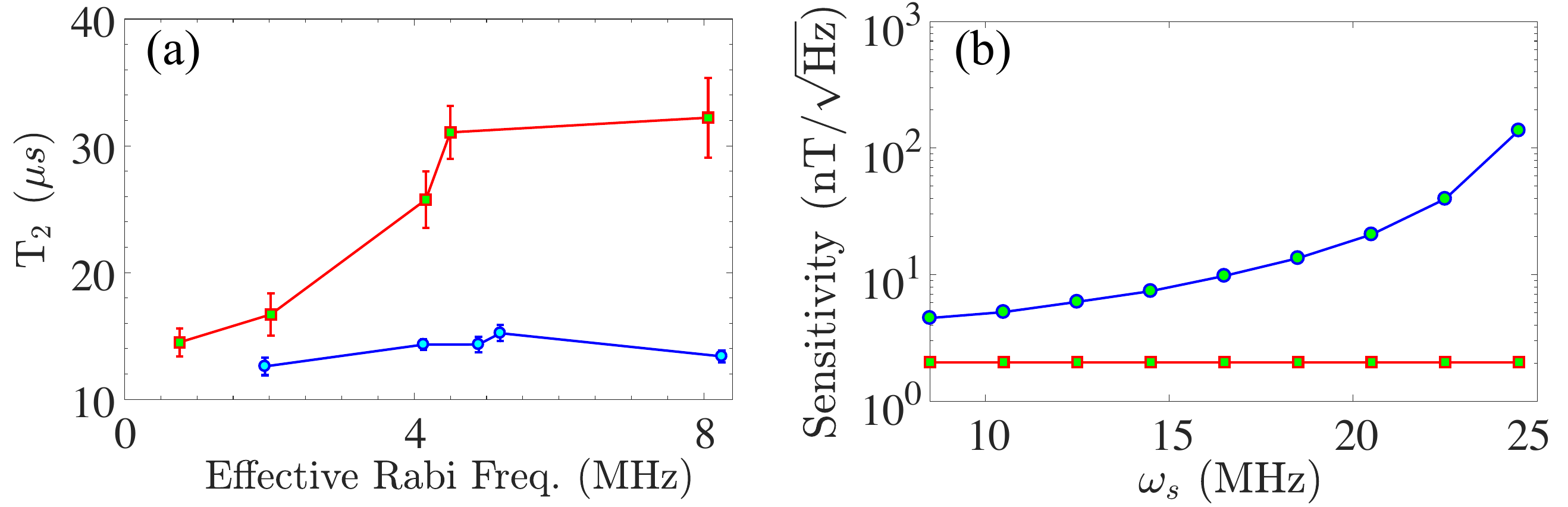}
\end{minipage}
\caption{\textbf{(a)} The extended coherence times as achieved by CCDD scheme (\protect\scalebox{0.6}{\color{Red}$\square$}) and XY8 pulse dynamical decoupling ({\color{Blue}$\circ$}) as a function of the effective Rabi frequency. The values in panel are the number of XY8 cycles (up to 12) that achieve the longest possible coherence time. \textbf{(b)} The estimated sensitivity for the measurement of an oscillating field using CCDD (\protect\scalebox{0.6}{\color{Red}$\square$}) and XY8 ({\color{Blue}$\circ$}) scheme. The achievable Rabi frequency is $\Omega=8.5$MHz, the coherence time is assumed as $T_2=15.22 \mu s$ (XY8) and $T_c=31.22 \mu s$ (CCDD). The oscillating field strength is $\gamma b=(2\pi)100$kHz. The other experiment parameters are the same as Fig.\ref{fig:XY8_CCD}.}
\label{fig:XY8_CCD_sen}
\end{figure}
%

Apart from avoiding the damaging effect on biological tissues (which is mainly dependent on the average microwave power), the constraint of microwave power on pulsed dynamical decoupling efficiency when combining with ({\it in vivo}) quantum sensing, may arise from another origin. The maximal microwave power, which determines the achievable Rabi frequency, may be limited due to e.g. the relatively large distance between microwave antenna and nanodiamond especially when compared to bulk diamond experiments. We perform measurements using different peak microwave power and compare with XY8 pulsed dynamical decoupling, the number of which achieves the longest possible coherence time. Our result as shown in Fig.\ref{fig:XY8_CCD_sen}(a) demonstrates that given the same effective Rabi frequency, i.e. the same average microwave power, the CCDD scheme clearly outperforms the XY8 dynamical decoupling sequences in extending spin coherence time. We remark that, unlike XY8 pulsed scheme where the decoupling efficiency in our experiments is limited by the constraint of microwave power, the achieved coherence time by CCDD is predominantly limited by the $T_1$ time of nanodiamonds, which can be improved by material design.

\section{Application in quantum sensing}

For sensing applications inside living cells, one has to take the constraint of microwave power arising from both origins discussed above into account. We consider a typical scenario of detecting an oscillating magnetic field with a frequency $\omega_s$. We remark that the underlying principle is essentially similar to the detection of electron (nuclear) spin, where the characteristic frequency is Larmor frequency of the target spins. One potential interesting example is the detection of the emergence or disappearence of radicals and functional molecule groups of nuclear spins. Pulsed scheme detects the field by engineering the time interval $\tau_p$ between pulses to match the field frequency, namely $\tau_c = k (\pi/\omega_s)$. For ideal instantaneous $\pi$-pulses (requiring infinite microwave power), the estimated measurement sensitivity is $\eta_c=k\pi/(4\gamma\sqrt{T_2})$, where $\gamma$ is the electronic gyromagnetic ratio and for simplicity we assume a unit detection efficiency. However, the limited achievable pulse peak Rabi frequency $\Omega$ (in comparison with the field frequency $\omega_s$) leads a constraint on the resonant condition $k\ge (\omega_s/\Omega)$ and would decrease the signal contrast. This fact restricts pulsed schemes to work only for frequencies below $\sim$ 10 MHz when the pulse repetition rate and microwave power are limited which is quite likely in biological systems. The present CCDD scheme can detect the field on resonance when $\Omega_1=\omega_s$  and $\omega_0-\omega_s=\pm \Omega_1$ with the estimated measurement sensitivity $\eta_c=1/(\gamma \sqrt{T_c})$ and $2/(\gamma \sqrt{T_c})$ respectively \cite{Cohen16}. We remark that the resonance condition can be satisfied by tuning $\omega_0$ (via the external magnetic field), thus it is flexible to choose $\Omega_1$ and $\Omega_2$ following the principle to optimise the improvement of coherent time. Therefore, the advantage of CCDD scheme as compared with the pulsed scheme in quantum sensing is not only the improvement of coherence time, but also its capability to increase sensitivity in the presence of limited average microwave power constraint. In Fig.\ref{fig:XY8_CCD_sen}(b), we compare the estimated measurement sensitivity for CCDD and the pulsed scheme under the same average power constraint. It can be seen that CCDD shows superior performance for signal frequencies above $10$ MHz. Potentially interesting examples include but not limited to the detection of molecules (containing nuclear spins) in high field magnetic resonance spectroscopy and electron (radicals). Besides the sensitivity enhancement, we also remark that CCDD scheme may avoid the misidentification of frequency components in classical fields or single molecule spectroscopy \cite{Loretz15,Haase16,Shu2017} due to the relatively long pulse duration (as microwave power is not sufficiently high). The present scheme (with both the sensitivity and the linewidth limited by the extended $T_2$) would thus advance the application of N-$V$ based quantum sensing in vivo using nanodiamonds, offering new methodology complementary to continuous wave ESR resonance measurement (the sensitivity is limited by the short $T_2^*$) and relaxation spectroscopy (the linewidth is limited by $T_2^*$) \cite{Hall16}.

\section{Conclusion and discussion}

To conclude, we implement a concatenated continuous dynamical decoupling strategy to prolong the quantum spin coherence time of N-$V$ centers in nanodiamonds even inside living cells. We demonstrate significantly increased performance with less stringent requirement on microwave power in comparison with pulsed schemes, and thus causing less severe damage effect to living cells. The concatenated continuous dynamical decoupling strategy thus provides a valuable tool to achieving long spin coherence time quantum sensors when the available and feasible microwave power is low or has to be limited, e.g. to avoid damage to biological tissue. It also enables relatively high-frequency magnetic field sensing with a substantial enhancement in the measurement sensitivity. The ability to extend spin coherence times in living cells along with enhanced measurement sensitivity raises new possibilities for the applications of nano diamond based quantum sensing in the intra-cellular environment and related biological events.

\section{Acknowledgements}

We thank Jianwei Wang, Quan Gan, Yuzhou Wu and Yuan Zhuang for help in sample preparation, Michael Ferner, Manfred B{\"u}rzele, Z.-J. Shu, J.-Y. He, R.-F. Hu and H.-B. Liu for technical assistance and Itsik Cohen for fruitful discussion. We acknowledge support by National Natural Science Foundation of China (11874024, 11690032), and the Open Project Program of Wuhan National Laboratory for Optoelectronics  NO. 2019WNLOKF002. A.R acknowledges the support of the ERC grant QRES. M. B. P. is supported by the DFG (FOR1493), the EU via DIADEMS and HYPERDIAMOND, the ERC Synergy grant BioQ and the IQST. C. M., N. T., B. N., L. P. M., F. J. are supported by DFG (FOR 1493, SFB TR21, SPP 1923), VW Stiftung, BMBF, ERC, EU (DIADEMS), BW Stiftung, Ministry of Science and Arts, Center for Integrated quantum science and technology (IQST).

\section*{Appendix}

\subsection{Principle of concatenated continuous dynamical decoupling}

\renewcommand{\theequation}{A\arabic{equation}}
\setcounter{equation}{0}

To suppress the effect from both environment noise and microwave fluctuation, we implement concatenated continuous dynamical decoupling by introducing a microwave driving field with time-dependent phase modulation \cite{Cohen16}. We repeat here the derivation for self consistency. We start with the Hamiltonian:
\begin{equation}
H=\frac{\omega_{0}}{2}\sigma_{z} + (\Omega_{1} + \delta_x)\cos\blb{\omega_{0}t+2\frac{\Omega_{2}}{\Omega_{1}}\sin\bla{\Omega_{1}t}}\sigma_{x}
\end{equation}
By moving to the interaction picture with respect to:
\begin{eqnarray}
H =\blb{ \frac{\omega_{0}}{2} + \Omega_{2}\cos(\Omega_{1}t) }\sigma_{z}
\end{eqnarray}
we get:
\begin{equation}
H_1 =\bla{ \frac{ \Omega_1 + \delta_x }{2}} \sigma_x - \Omega_2\cos(\Omega_{1}t) \sigma_{z}
\end{equation}
Moving again to the interaction picture with respect to $\Omega_1 \sigma_x$ we get:
\begin{equation}
H_2 = \delta_x \sigma_x - \bla{\Omega_2/2} \sigma_z
\end{equation}
as $\Omega_2$ is generated by the time dependent phase we assume that the noise is negligible.
As $\Omega_2 \gg \delta_x$ and we assume that the amount of $\Omega_2$ in the power spectrum of $\delta_x$ is negligible the deleterious effect of the noise will only manifest itself in second order, i.e.,
\begin{equation}
H_{noise} =\bla{ {\delta_x^2}/{\Omega_2}} \sigma_z
\end{equation}
as $\bla{\Omega_2/\Omega_1}$ is kept at $10^{-1}$ and $\delta_x$ is of the order of $1\%$ of $\Omega_1$ the effect of the noise is of the order of $10^{-3} \Omega_1.$
It is noteworthy that this effect could be further suppressed by adding a higher drive by an extra time dependent phase term. In our experiment, we first apply a $(\pi/2)_y$ pulse to prepare the N-$V$ centre spin in a superposition state $\ket{\psi(0)}=(1/\sqrt{2})(\ket{0}+\ket{1})$. After an evolution for time $t$, the N-$V$ centre spin state evolves to
\begin{equation}
\ket{\psi(t)} = \exp\blb{(-it\Omega_1/2)\sigma_x}\exp\blb{(-it\Omega_2/2)\sigma_z}\ket{\psi(0)}.
\end{equation}
The fluorescence measurement after another $(\pi/2)_y$ pulse gives the state population
\begin{equation}
\mbox{P}_{\ket{0}} = |\langle \psi(0) \vert \psi(t)\rangle|^2=\frac{1}{2}\blb{1+\cos(\Omega_1 t)\cos(\Omega_2 t)}.
\end{equation}
Two frequency components $\Omega_1$ and $\Omega_2$, in additional to the effect of unpoloarized nitrogen nuclear spin, leads to the beating pattern in the oscillating signal, which explain our experiment observation and is also verified by numeric simulation. The extended coherence time can be inferred from the decay of the envelope. \\
\begin{table*}
\begin{tabular}{c|c|c|c|c|c|c|c|c}
 & 1 & 2 & 3 & 4 & 5 & 6 & 7 & 8\tabularnewline
\hline
$T_{2}$($\mu$s) & 4.840$\pm$0.009 & 4.300$\pm$0.069 & 2.142$\pm$0.018 & 1.30$\pm$0.24 & 4.292$\pm$0.133 & 8.710$\pm$0.203 & 2.990$\pm$0.083 & 2.711$\pm$0.078\tabularnewline
\hline
$\alpha$ & 1.082$\pm$0.045 & 1.542$\pm$0.056 & 1.448$\pm$0.026 & 1.447$\pm$0.295 & 1.47$\pm$0.10 & 1.392$\pm$0.068 & 1.576$\pm$0.101 & 1.740$\pm$0.113\tabularnewline
\end{tabular}
\caption{{\bf Spin echo measurement for nanodiamond N-$V$s.} The normalized experiment data is fitted with the function $P_{\ket{0}}(t)=(1/2)[1+e^{-(t/T_{SE})^{\alpha}} ]$. The table lists the estimated value of $T_{SE}$ and $\alpha$ for 8 nanodimonad N-$V$s.}
\label{si-fig:spinecho}
\end{table*}

\subsection{Characteristics of nanodiamonds}

\renewcommand{\theequation}{B\arabic{equation}}
\setcounter{equation}{0}

The nanodiamonds are spin-coated on the mica plate and scanned after drying, we choose different areas to perform AFM scan and count the statistics of nanodiamond size. Considering the broadening effect of AFM, we measure the height of nanodiamonds to estimate their sizes. The statistics of nanodiamond size is shown in Fig.1(b) in the main text. The average nanodiamond size is about $43$ nm, with the diameters predominantly within $25-60$ nm ($43\pm 18$ nm), and the nanodiamond concentration is about $50$ per $(10 \mu m)^{2}$.
%

%
\begin{figure}[h]
\begin{minipage}{9cm}
\hspace{-0.3cm}
\includegraphics[width=6.5cm]{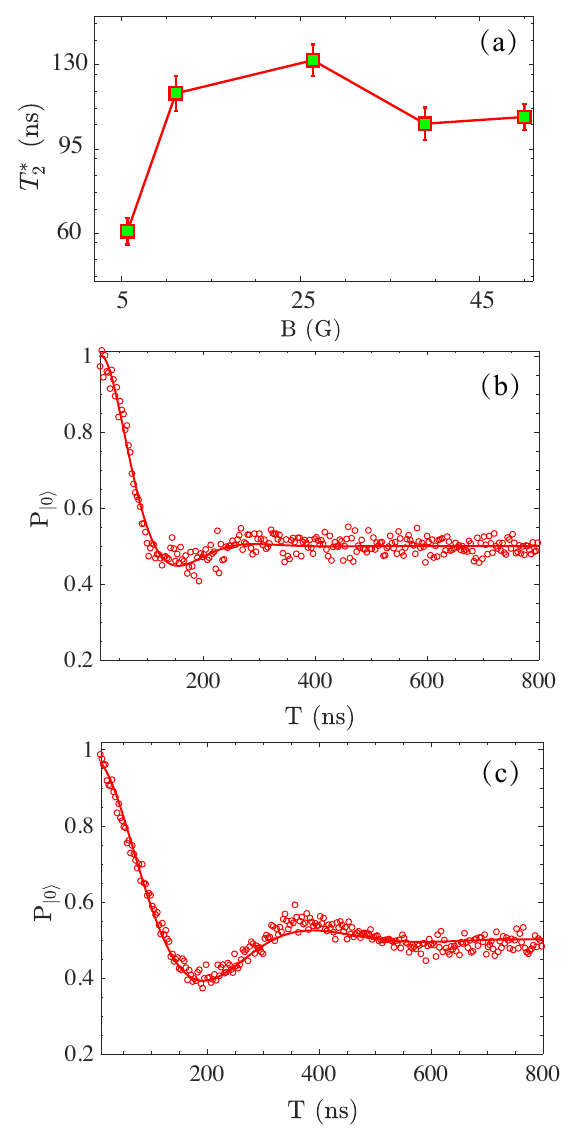}
\end{minipage}
\caption{{\bf (a)}  Dependence of $T_{2}^{*}$ on the strength of magnetic field. {\bf (b)} and {\bf (c)} show the data of Ramsey measurement with magnetic field $B=5$ G {\bf (b)} and $B=25$ G {\bf (c)}.  }\label{si-fig:ramsey}
\end{figure}
%

%
\begin{figure}[h]
\begin{minipage}{8cm}
\hspace{-0.5cm}
\includegraphics[width=8.5cm]{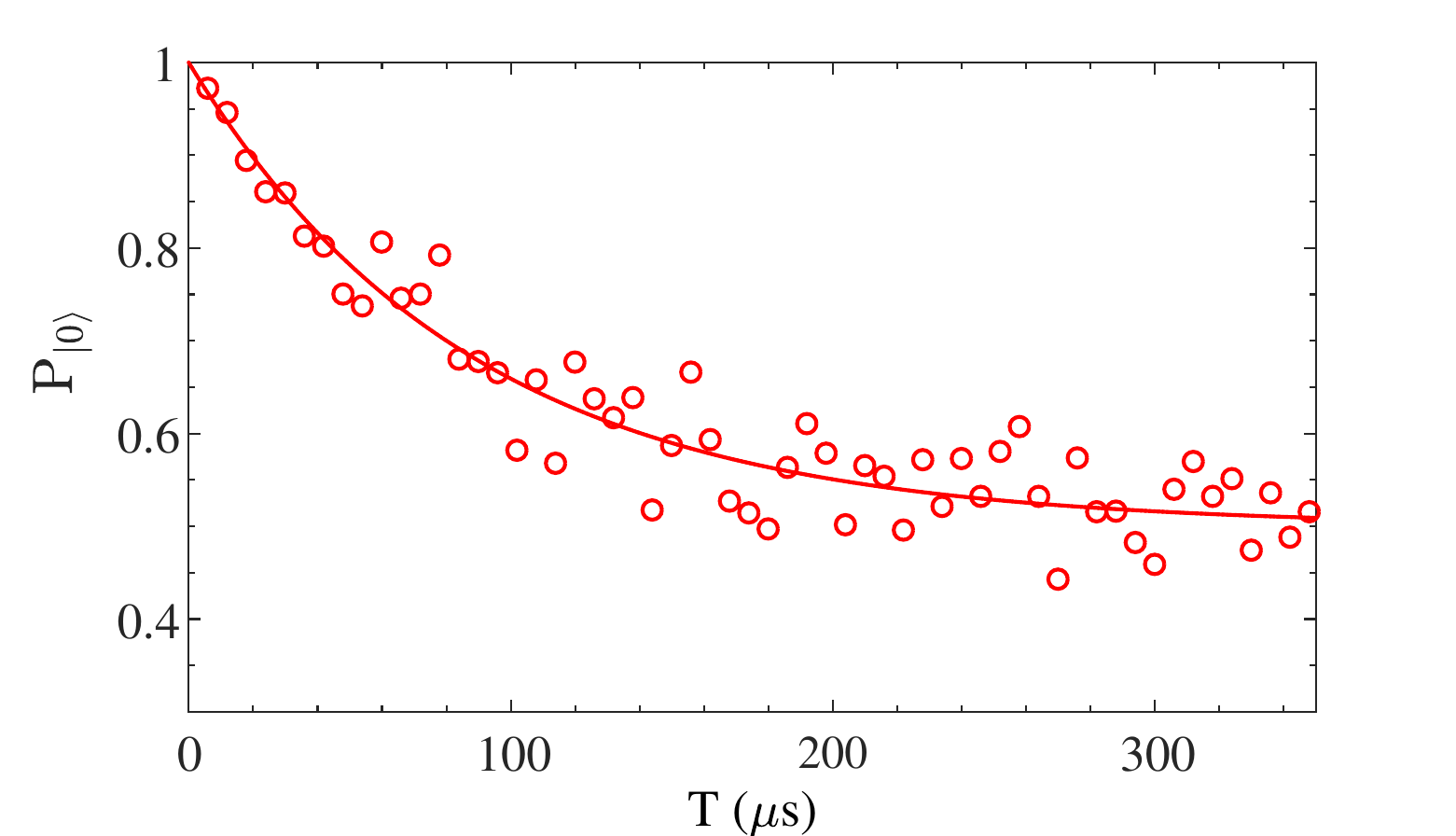}
\end{minipage}
\caption{The relaxation time of nanodiamond N-$V$. The normalized experiment data is fitted with the function $P_{\ket{0}}(t)=(1/2)[1+e^{-(t/T_{1})} ]$. The applied magnetic field is $B_{\parallel}=508$ Gauss.The relaxation time is estimated to be $T_1=87.35 \pm7.50 \mu s$.}
\label{si-fig:T1}
\end{figure}
%

To characterize the properties of spin noise, we perform Ramsey measurement applying magnetic field of different strength along N-$V$ axis. The dependence of $T_{2}^{*}$ on the applied magnetic field is shown in Fig.\ref{si-fig:ramsey}. It can be seem that $T_{2}^{*}$ increases with a larger magnetic field, which suggests that electric noise may be a dominant source of spin dephasing in the present sample \cite{Jac16}. We also perform spin echo measurement of N-$V$s in several nanodiamonds. Owing to the different environment for different nanodiamond N-$V$s, they exhibit different coherence times. The experiment data is normalized through Rabi oscillation, and then is fitted with the function $P_{\ket{0}}(t)=(1/2)[1+\exp\bla{-(t/T_{SE})^{\alpha}} ]$. The result shows that $\alpha\in[1.08,1.74]$ and $T_{SE}$ for the average spin echo coherence time is around $3\mu s$, see Table \ref{si-fig:spinecho}.
In Fig.\ref{si-fig:T1}, we plot the experiment data for the relaxation time measurement of the N-$V$ center shown in Fig.2 of the main text. The normalized experiment data is fitted with the function $P_{\ket{0}}(t)=(1/2)[1+\exp\bla{-(t/T_{1})} ]$, which gives an estimation of $T_1=87.35 \pm7.50 \mu s$.\\

In the main text, we present experiments in which we apply XY8-N pulse sequences to extend coherence time of N-$V$ centres using a train of $(8\times N)$ $\pi$-pulses. The extended coherence time $T_2$ under pulsed dynamical decoupling increases as the number of XY8 cycles grows, and reaches a saturates value. In Fig.4(a) of the main text, we apply $\pi$-pulses with different peak Rabi frequency and measure the achievable saturate coherence time. The average microwave power  can be quantified by the average effective Rabi frequency which is defined as follows
\begin{equation}
\bar{\Omega}=\blb{\frac{1}{T}\int_{0}^{T} \Omega^2(t)dt}^{1/2}
\end{equation}
In Table~\ref{si-fig:xy8}, we list the peak Rabi frequency, the average effective Rabi frequency and the number of XY8 cycles of the pulse sequences that achieve the best coherence time as shown in Fig.4(a) of the main text. \\

\begin{table*}
\begin{tabular}{l|c|c|c|c|c}
 & 1 & 2 & 3 & 4 & 5  \tabularnewline
\hline
Peak Rabi frequency (MHz) & $\quad$1.98 $\quad$ & $\quad$ 5.10$\quad$ &$\quad$ 7.14 $\quad$& $\quad$8.50$\quad$ & $\quad$16.20  $\quad$ \tabularnewline
\hline
Average effective Rabi frequency (MHz)& 1.94 & 4.13 & 4.89 & 5.19 & 8.25   \tabularnewline\hline
XY8 cycles & 6 & 12 & 12 & 12 & 14   \tabularnewline
\end{tabular}
\caption{The detailed information on the XY8 pulse sequences that achieve the coherence time of N-$V$s in nanodiamond as shown in Fig.4(a) of the main text.}
\label{si-fig:xy8}
\end{table*}
%

%
\begin{figure*}
\begin{minipage}{15cm}
\hspace{-0.2cm}
\includegraphics[width=15cm]{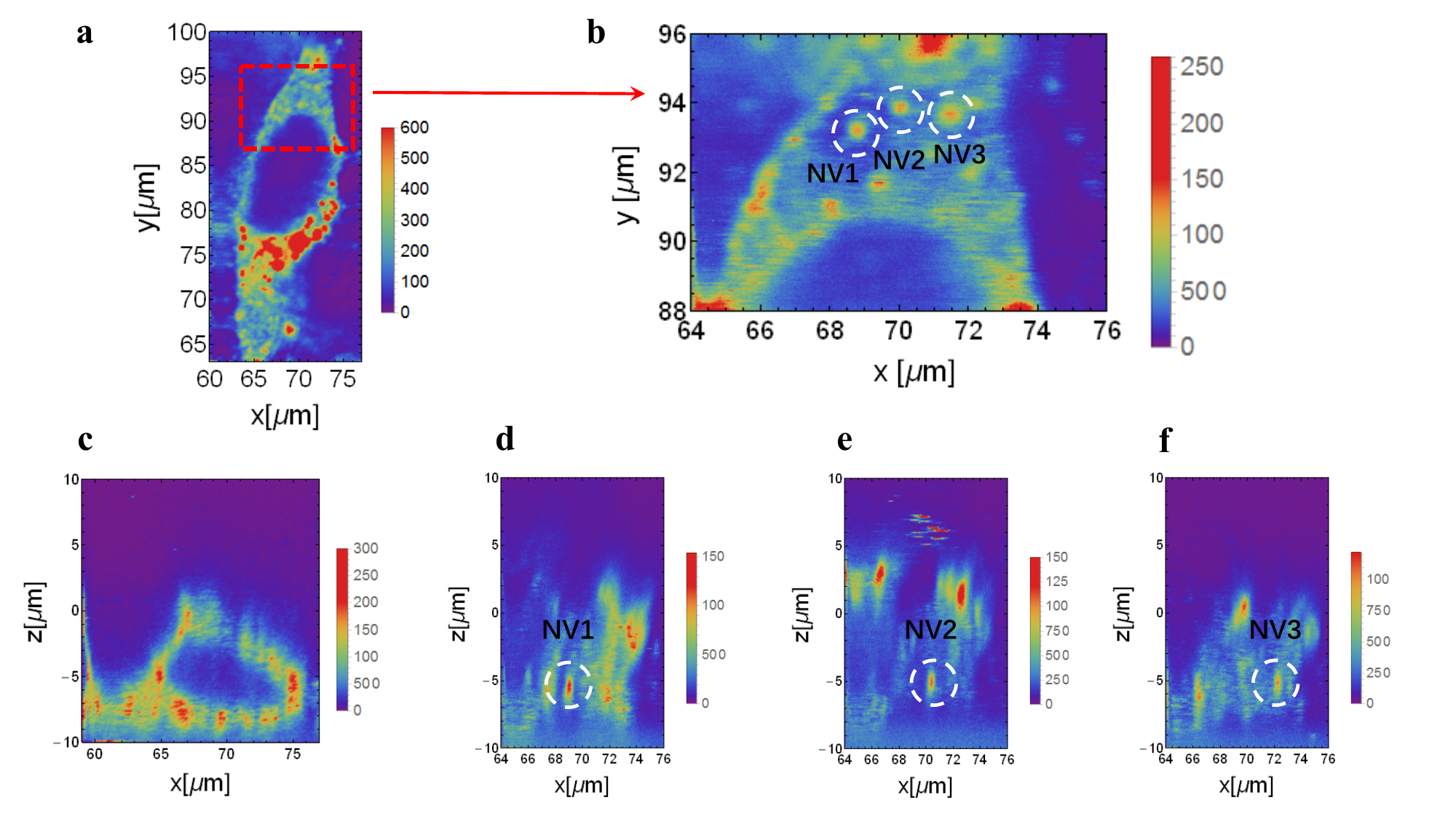}
\end{minipage}
\caption{{\bf (a)} Confocal scan of a living cell with cytomembrane lipophilic fluorescent dyes. The zoom in area in the dashed rectangular is shown in {\bf (b)}. Three nanodiamonds are identified in the cell, which is confirmed by XZ scan as shown in {\bf (c-f)}. The photon count is in the unit of kcs/s.}
\label{si-fig:cell}
\end{figure*}

\subsection{Cell culture and sample preparation}

\renewcommand{\theequation}{C\arabic{equation}}
\setcounter{equation}{0}

NIH/3T3 cells were cultured in Dulbecco's modified Eagle's medium (DMEM) supplemented with fetal bovine serum and penicillin/streptomycin. Nanodiamonds were diluted in DMEM. Then cells were seeded on cover slips and incubated with the DMEM-nanodiamond suspension (37\textdegree{}C, 5\% CO$_{2}$) for 20 hours, which allowed cells to adherent to the surface of the cover glass. After treatment, the media was removed and the cover glass was washed 3 times with phosphate buffered saline (PBS). The cultured NIH/3T3 cells were immersed in PBS throughout the measurement, and the temperature was kept around 22\textdegree{}C. The confocal imaging was performed through the cover glass, using an oil immersion lens. The N-$V$ center measured in this work was located about 2.5 micrometers above the cover glass. Quantum measurements were performed on the N-$V$ centers by applying a microwave signal along a copper wire (with a diameter $\sim$20$\mu$m) which is about $\sim$30$\mu$m far away from the N-$V$ center.\\

\subsection{Cell membrane staining and identifying nanodiamonds in cells}

\renewcommand{\theequation}{D\arabic{equation}}
\setcounter{equation}{0}

In order to identify those nanodiamonds that were taken in cells, we first use the 1, $1^{\prime}$-Dioctadecyl-3,3,$3^{\prime}$,$3^{\prime}$-tetramethylindocarbocyanine perchlorate(DiIC$_{18}(3)$) which is a kind of lipophilic fluorescent dyes to label the cytomembrane. After the procedure of cellular uptake, the cell culture medium was removed, and the cells were then incubated with 200 $\mu$L DiIC$_{18}(3)$ (0.0486 $\mu$M/L) solution for about 1 hour (37\textdegree{}C, 5\% CO$_{2}$).  The sample was washed 5 times with PBS before it was imaged with a home-built confocal setup. The stained cell images with clear profile are shown in Fig.\ref{si-fig:cell}(a,c), which demonstrate clearly the cytomembrane labelled by DiIC$_{18}(3)$ and the cell nucleus. Fig.\ref{si-fig:cell}(b) shows a zoom in area where we identify three nanodiamonds in cell, which is further confirmed by our XZ confocal scan, see Fig.\ref{si-fig:cell}(d-f).\\

\subsection{Influence of microwave radiation on living cells}

\renewcommand{\theequation}{E\arabic{equation}}
\setcounter{equation}{0}

\begin{figure}[b]
\begin{minipage}{9cm}
\hspace{-0.5cm}
\includegraphics[width=8.5cm]{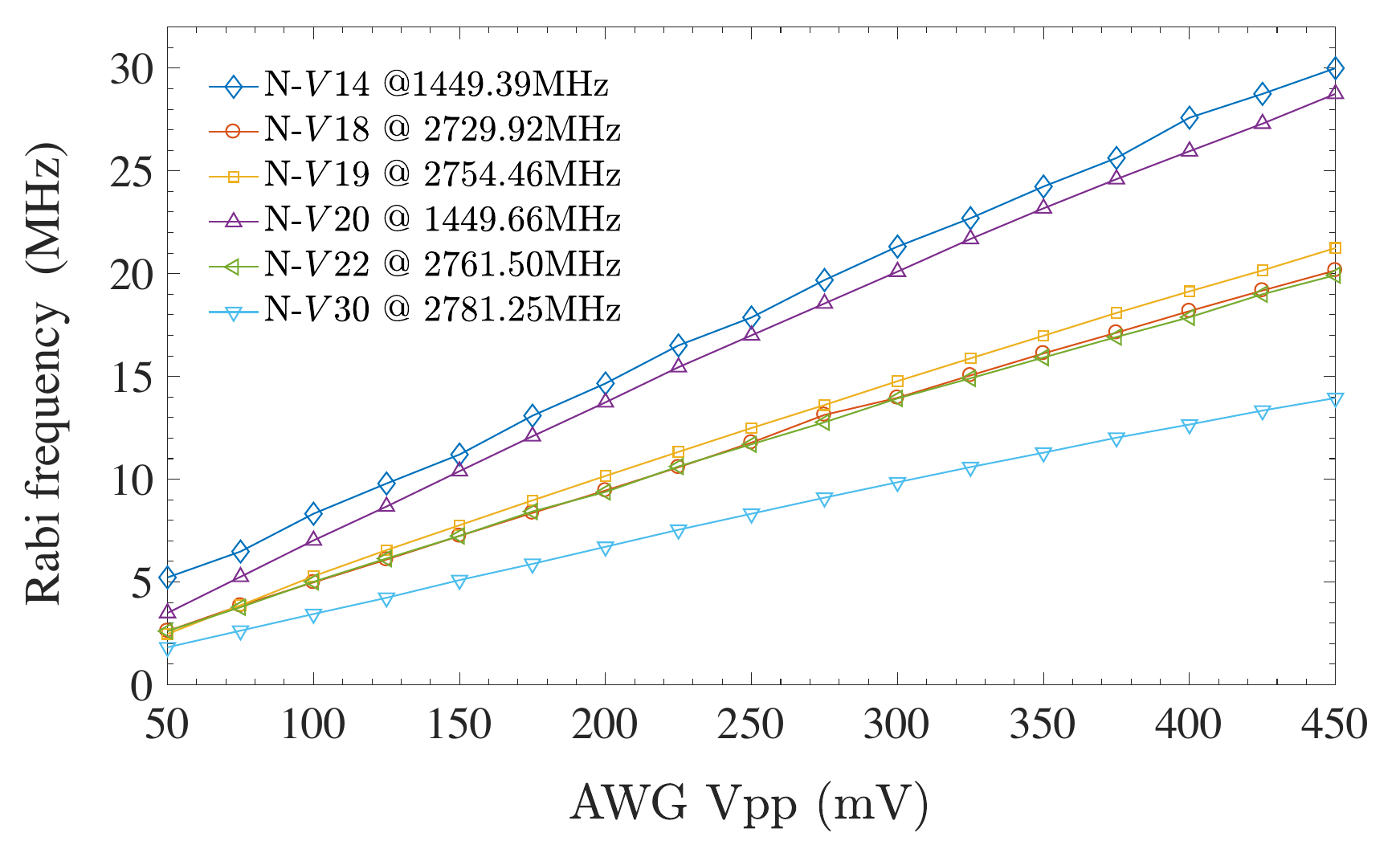}
\end{minipage}
\caption{Dependence of Rabi frequency on the peak-to-peak voltage of AWG output for several N-$V$s in nanodiamond. Rabi measurements are performed at frequencies as shown in the figure when applying different magnetic fields along the N-$V$ axis. The amplification percentage of the amplifier is fixed as $45\%$.}
\label{si-fig:rabi_apm}
\end{figure}

The main influence of microwave radiation on living cells arise from the heating effect. We perform measurements to clarify the following two issues: (1) The dependence of the heating effect on the microwave power; (2) Whether the difference in the required microwave power to achieve the same coherence times under pulsed and CCDD schemes matters, i.e. whether it may lead to non-negligible difference in the heating of the sample. The role of microwave power in the manipulation of N-$V$s is manifested by the observed Rabi frequency. In Fig.\ref{si-fig:rabi_apm}, we calibrate the dependence of Rabi frequency on the amplitude of AWG output by measuring Rabi frequencies for several N-$V$s in nanodiamond when applying magnetic fields of different strengths. The measurement shows that Rabi frequencies are proportional to the amplitude of the AWG output that is quantified by the peak-to-peak voltage. The results also provide us information on the requirement of microwave power to achieve a certain Rabi frequency. For example, Rabi frequency is promoted by 3-6 MHz with an increase of AWG output amplitude by 100mV for typical nanodiamonds that locate within a distance of $5-15\mu m$ to the microwave wire. For the measurement in Fig.4 of the main text, Rabi frequency of $9.6$ MHz (4.6 MHz) is achieved by an AWG output of 400 mV (200 mV) and $30\%$ ($25\%$) amplification percentage. As $\Omega_2 \ll \Omega_1$, one can easily verify that the microwave radiation power is determined by the amplitude $\Omega_1$.

%
%

It has been shown that the main damage of microwave (MW) radiation applied to cells is caused by the heating effect. The viability of cells may be influenced substantially as the temperature increases by 10$^{\circ}$C, see e.g. Ref.\cite{Woo2000,Shamis2012}. In our experiment, we apply microwave of pulse and CCDD schemes and monitor the temperature of sample under similar conditions in the experiments as presented in the main text. We glue a coverslip (2.4cm $\times$ 2.4cm), adhere to which cells grow, on a PCB board in contact with a copper wire that delivers microwave. The thermistor (Thorlabs TH10K) attached to the coverslip allows us measure the temperature of the sample when applying microwave sequences. We add 1mL phosphate buffered saline (PBS) which was used to maintain gentle conditions for living cells.

\begin{figure}[t]
\begin{minipage}{9cm}
\hspace{-0.2cm}
\includegraphics[width=8.5cm]{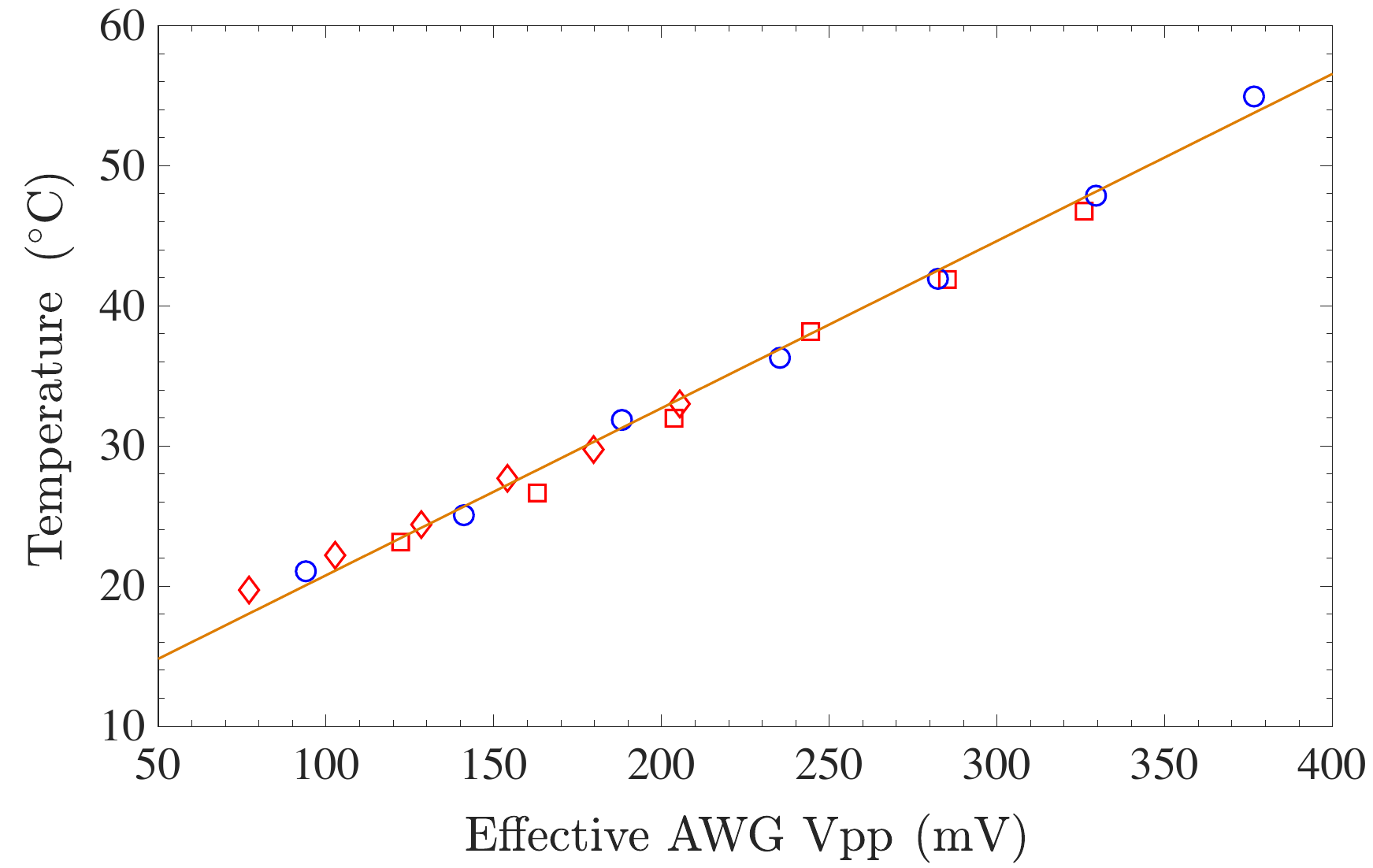}
\end{minipage}
\caption{The increase of sample temperature as a function of the average amplitude of AWG output. The amplification percentage of the amplifier is fixed as $45\%$.  The results for CCDD scheme are shown in {\color{Blue}$\circ$} as compared with the pulse scheme (\protect\scalebox{0.6}{\color{Red}$\square$} and \protect\scalebox{1}{\color{Red}$\diamond$}).}
\label{si-fig:tem_apm}
\end{figure}

We denote the peak-to-peak voltage of AWG output as $A$ (mV), which quantifies the peak power of microwave radiation. We first apply CCDD scheme with the following parameters: the duration of continuous microwave driving is $T_{\mbox{on}}=30 \mu s$ (which is similar to the extended coherence time as observed in our experiment), the combination of laser pulse and idle time is set as $T_{\mbox{off}}=3.8 \mu s$ (which is similar to the corresponding time in our experiments). As there is time during which microwave is switched off, we define the average amplitude of microwave radiation as follows
\begin{equation}
\langle A\rangle = A\sqrt{ \frac{ T_{\mbox{on}} }{T_{\mbox{on}}+T_{\mbox{off} } } }.\label{eq-si:avergae_amplitude_pulse}
\end{equation}
We monitor the temperature of the sample every 60 seconds and record the stable temperature which usually is reached in $\sim 15$ minutes. The result is shown in Fig.\ref{si-fig:tem_apm} ({\color{Blue}{$\circ$}}). For comparison, we also apply XY8-12 scheme (i.e. the total number of pulses is $N=96$) with the following parameters: the $\pi$-pulse duration is $\tau_{\pi}=50$ ns, and the time between pulses (i.e. no microwave) is $\tau_{\mbox{f}}=100 (4)$ ns, the combination of laser pulse and idle time is set as $T_{\mbox{off}}=3.8 \mu s$. In this case, the average amplitude of microwave radiation is given by
\begin{equation}
\langle A\rangle = A\sqrt{ \frac{ N\tau_{\pi} }{N\tau_{\pi}+N\tau_{\mbox{f}}+T_{\mbox{off} } } }.\label{eq-si:avergae_amplitude_ccd}
\end{equation}
The results are shown in Fig.\ref{si-fig:tem_apm} (\protect\scalebox{0.6}{\color{Red}$\square$} for $\tau_{\mbox{f}}=100$ ns and \protect\scalebox{1}{\color{Red}$\diamond$} for $\tau_{\mbox{f}}=4$ ns). Although we choose certain specific parameters of pulses (which are close to those parameters used in our experiments), we find that the heating effect is mainly determined by the {\it average} microwave power (as quantified by the average amplitude of AWG output defined in Eq.\ref{eq-si:avergae_amplitude_pulse}
-\ref{eq-si:avergae_amplitude_ccd}) for both pulse and CCDD schemes while is weakly dependent on the details of microwave pulses. The sample temperature would heat up by $\sim 12 ^\circ$C when the average amplitude of AWG increases by 100 mV (which corresponds to $\sim$3-5 MHz improvement in Rabi frequency). We remark that the exact difference in heating effect may also depend on the other factors such as the distance between the microwave wire and the orientation of the N-$V$ axis, nevertheless, our measurement strongly suggests that the more stringent requirement on microwave power by pulse schemes will cause more severe heating effect to biological tissues. For example, we compare the data in Fig.4 of the main text. The average microwave power for CCDD to achieve $\sim 15 \mu s$ coherence time is less than $1$ MHz, while it requires $4-5$ MHz for pulse scheme. Therefore, according to the observed characteristics of heating effect due to microwave radiation, the temperature increase due to the pulsed sequences would be about 10~$^{\circ}$C more than that due to CCDD, which can be expected to have a significant effect on biological tissues \cite{Woo2000,Shamis2012}.

%
%
%

%
\subsection{Sensitivity comparison under the constraint of microwave power}

\renewcommand{\theequation}{F\arabic{equation}}
\setcounter{equation}{0}

We consider a typical scenario of quantum sensing, namely the detection of a weak oscillating magnetic field $b(t) = b \cos(\omega_s t)$ with a frequency $\omega_s$ and an amplitude $b$. The underlying principle is similar to the detection of electron (nuclear) spin in essence, where the characteristic frequency is Larmor frequency of the target spins. One possible interesting example is the detection of radicals inside cells. In N-$V$ spin sensor based magnetic spectroscopy, the Larmor frequency of the target spins would usually exceed $\sim 10$ MHz, e.g. for electron spin or nuclear spin in high-field magnetic resonance spectroscopy. Depending on the relative orientation of the magnetic field with respect to the N-$V$ axis (denoted as the $\hat{z}$ axis that connects the nitrogen atom and the vacancy site) and the field frequency $\omega_s$, the N-$V$ centre spin sensor may be sensitive to either the field components along the $\hat{z}$ direction or the $\hat{x}$ direction. In order to investigate the achievable measurement sensitivity under the constraint of microwave power, we denote the available (maximum) Rabi frequency as $\Omega_{max}$. \\

%
\begin{figure}[b]
\begin{minipage}{9cm}
\hspace{-0.2cm}
\includegraphics[width=8cm]{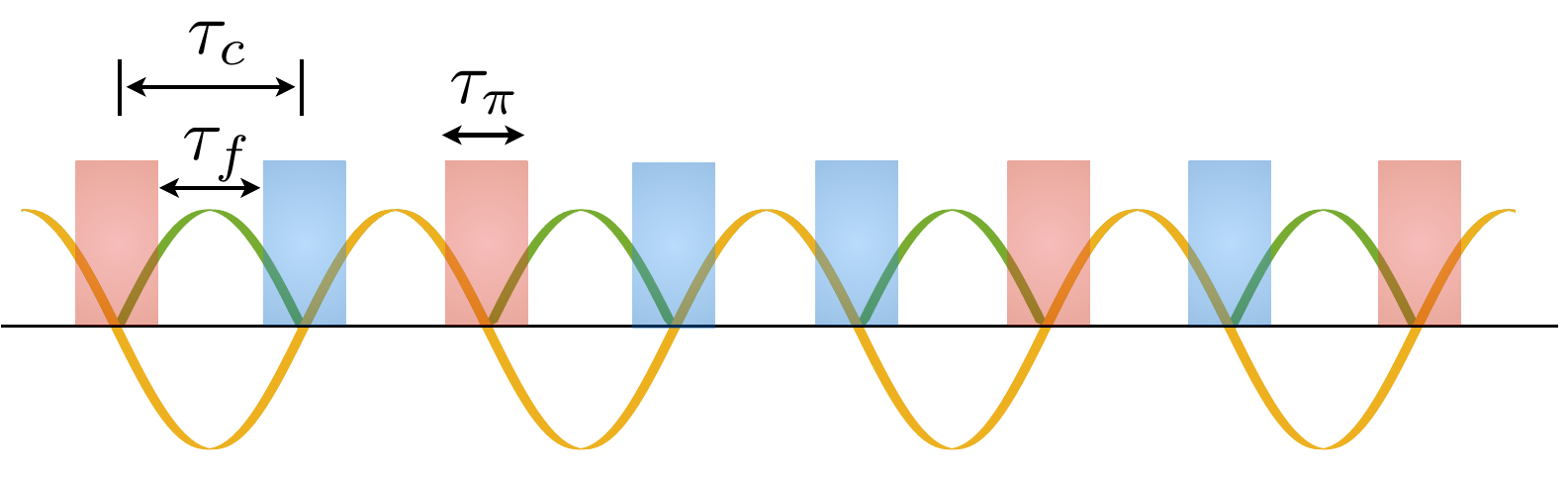}
\end{minipage}
\caption{Pulse scheme with non-instantaneous $\pi$-pulses. The $\pi$-pulse duration is denoted as $\tau_{\pi}=\bla{\pi/\Omega}$ where $\Omega$ is Rabi frequency of pulses. The inter-pulse free evolution time is $\tau_f$. The time interval $\tau_c=\tau_{\pi}+\tau_f$ between pulses shall match the oscillating field frequency, namely $\tau_c = k (\pi/\omega_s)$, where $k=1,3,5,\cdots$. The role of $\pi$-pulses is to change the sign of the magnetic field that acting on the N-$V$ centre spin, so that its effect can be accumulated constructively.}
\label{si-fig:pulse_time}
\end{figure}
%

%
\begin{figure*}
\begin{minipage}{16cm}
\includegraphics[width=16cm]{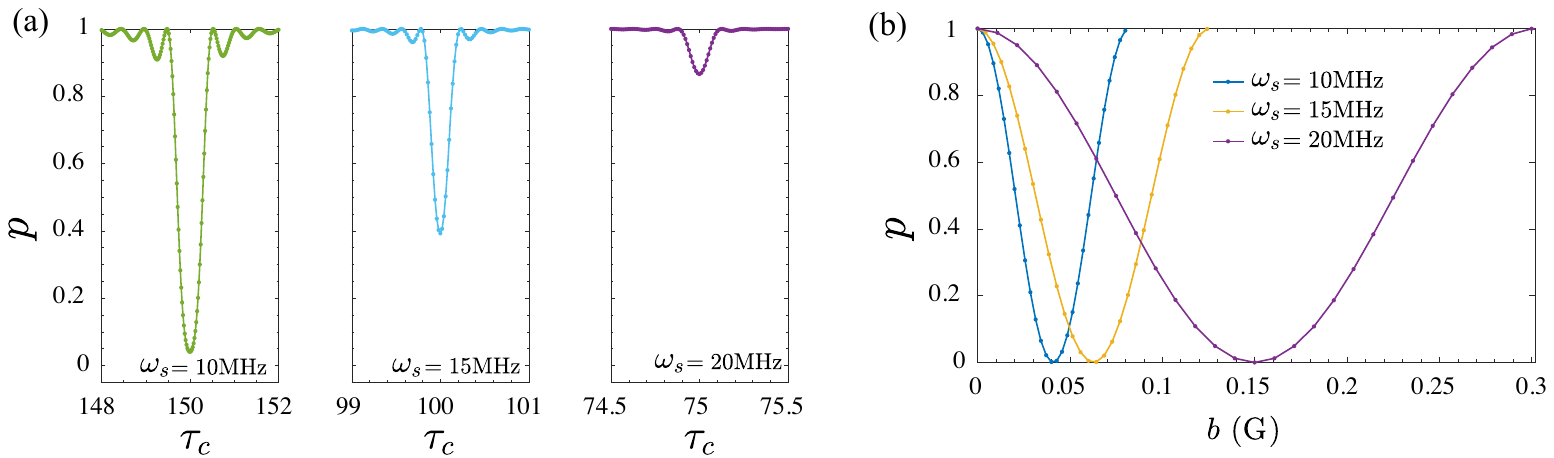}
\end{minipage}
\caption{Quantum sensing using pulse scheme with non-instantaneous $\pi$-pulses. (a) The signal as a function of  the time interval close to the resonant condition $\tau_c = 3 (\pi/\omega_s)$ for an oscillating magnetic field with different frequencies $\omega_s=10 $ MHz, $15$ MHz, $20$ MHz. The field strength is $(\gamma b)=(2 \pi) 100$ kHz. (b) The signal $p$ as a function of the oscillating field strength $b$. In (a-b), the interrogation time is set as $T_2=15.22\mu s$. The pulse Rabi frequency is $\Omega=8.5$ MHz.}
\label{si-fig:pulse_time_sens}
\end{figure*}

We first consider quantum sensing in combination with the present CCDD scheme. In the first case of $\omega_s \leq \Omega_{max}$, we can choose to measure the field along the $\hat{z}$ direction by tuning the orientation of the N-$V$ axis in parallel with the polarisation of the magnetic field. In the interaction picture, when setting $\Omega_1=\omega_s$, the effective Hamiltonian becomes
\begin{equation}
H_2=(\Omega_2/2) \sigma_z +(\gamma b/2)\sigma_z,
\end{equation}
where $\gamma$ is the electronic gyromagnetic ratio \cite{Cohen16}. The field parameter $b$ can be determined via a Ramsey experiment, and the measurement sensitivity with an interrogation time $t=T_c$ is estimated to be
\begin{equation}
\eta_c=\frac{\sqrt{\Delta^2 p}}{C(\partial p/\partial b)\sqrt{1/T_c}}  \simeq 1/ \bla{\gamma C \sqrt{T_c}},
\end{equation}
where $T_c$ is the extended coherence time, $p=(1/2)\blb{1+\cos(\gamma b T_c)}$ is the result of Ramsey experiment, $C$ represents the detection efficiency \cite{Tay_08_NP}. As $\omega_s\leq\Omega_{max}$, it is always feasible to choose Rabi frequency $\Omega_1=\omega_s$. In the second case of $\omega_s >\Omega_{max}$, we choose to measure the field along the $\hat{x}$ direction by tuning the orientation of the N-$V$ axis perpendicular to the polarisation of the magnetic field. In the interaction picture, when setting $\Omega_1=\omega_0-\omega_s$, the effective Hamiltonian becomes \cite{Cohen16}
\begin{equation}
H_3=-(\gamma b/4) \sigma_z .
\end{equation}
The measurement sensitivity via a Ramsey experiment with an interrogation time $t=T_c$ is estimated to be
\begin{equation}
\eta_c =\frac{\sqrt{\Delta^2 p}}{C(\partial p/\partial b)\sqrt{1/T_c}}  \simeq 2/ \bla{\gamma C \sqrt{T_c}},
\end{equation}
where the signal of Ramsey experiment is $p=(1/2)\blb{1+\cos(\gamma b T_c/2)}$.
Pulse scheme detects an oscillating field by engineering the time interval $\tau_c$ between pulses (see Fig.\ref{si-fig:pulse_time}) to match the field frequency, namely $\tau_c = k (\pi/\omega_s)$ \cite{Degen17}. For ideal instantaneous $\pi$-pulses (i.e. requiring infinite microwave power), $\tau_c=\tau_f$ where $\tau_f$ is the free evolution between pulses. In our experiment, we first apply a $(\pi/2)_{y}$ pulse to prepare the N-$V$ centre spin in a superposition state $\ket{\psi(0)}=(1/\sqrt{2})(\ket{0}+\ket{1})$. After an interrogation time for time $t$, the N-$V$ centre spin state evolves to the following state as
\begin{equation}
\ket{\psi(t)} = \exp\blb{-i \gamma b ( {2/\pi} )  t\sigma_z}\ket{\psi(0)}.
\end{equation}
The factor $( {2/\pi} ) $ comes from the average of the modulated oscillating field. The state population measurement after another $(\pi/2)_{y}$ pulse leads to the signal of Ramsey experiment as follows
\begin{equation}
p = |\langle \psi(0) \vert \psi(t)\rangle|^2=\frac{1}{2}+\frac{1}{2}\cos\blb{\gamma b ( {4/\pi} ) t }.
\end{equation}
The estimated measurement sensitivity with an interrogation time $T_2$ is
\begin{equation}
\eta_p=\frac{\sqrt{\Delta^2 p}}{C(\partial p/\partial b)\sqrt{1/T_2}}  \simeq k\pi/(4\gamma C\sqrt{T_2}).
\end{equation}
Given a field with a frequency $\omega_s$, the required power is $\Omega^2$ for the CCDD scheme with $\Omega=\omega_s$. For the pulsed scheme, the pulse repetition rate shall be $\pi/\omega_s$, namely the time interval between two pulses is $\tau_c = k (\pi/\omega_s)$. The pulse duration $\tau_{\pi}$ should be much smaller than $\tau_c$ (see Fig.\ref{si-fig:pulse_time} in supplementary information), namely the ratio $a=\tau_{\pi}/\tau_{c} \ll 1$. Thus, one can estimate that the required power for the pulsed scheme is $\Omega^2/a$. It can be seen that the power required by the CCDD scheme is less than the pulsed scheme by a factor of $a=\tau_{\pi}/\tau_{c} \ll 1$. \\

For non-instantaneous $\pi$-pulses realised by finite microwave power, $\tau_p=\tau_{\pi}+\tau_f$, where $\tau_{\pi}=\pi/\Omega$ is the pulse duration and $\tau_f$ is the free evolution between pulses, see Fig.\ref{si-fig:pulse_time}. As we are considering pulse scheme, we require $\tau_f>0$, otherwise the pulse scheme would become continuous. However, the limited pulse Rabi frequency $\Omega$ (in comparison with the field frequency $\omega_s$) leads to a constraint on the resonant condition $k\ge (\omega_s/\Omega)$ and would decrease the signal contrast. In Fig.\ref{si-fig:pulse_time_sens}, we shown the signal contrast for the measurement of an oscillating magnetic field with different frequencies using pulses of the same duration (namely the same available Rabi frequency). It can be seen that the signal contrast decreases with an increasing field frequency, see Fig.\ref{si-fig:pulse_time_sens}(a). This is also confirmed by the less steep signal slope $\bla{\partial p/\partial b}$ for an oscillating field with a higher frequency $\omega_s$, see Fig.\ref{si-fig:pulse_time_sens}(b). In Fig.4 of the main text, we compare the estimated measurement sensitivity by CCDD and pulse scheme according to the achieved coherence time. The enhanced sensitivity by the present CCDD scheme arises from both the prolonged coherence time and also the limit of microwave power constraint in pulse scheme.

\bibliographystyle{ieeetr}
\bibliography{99}

\begin{thebibliography}{60}
\bibitem{Doherty13} M. W. Doherty, N. B. Manson, P. Delaney, F. Jelezko, J. Wrachtrup, L. C. L. Hollenberg, {\it The nitrogen-vacancy colour centre in diamond}, \href{http://www.sciencedirect.com/science/article/pii/S0370157313000562}{{\it Phys. Reports} \textbf{528}, 1 (2013)}.

\bibitem{Maz_08_NL} J. R. Maze, P. L. Stanwix, J. S. Hodges, S. Hong, J. M. Taylor, P. Cappellaro, L. Jiang, M. V. G. Dutt, E. Togan, A. S. Zibrov, A. Yacoby, R. L. Walsworth, M. D. Lukin, {\it Nanoscale magnetic sensing with an individual electronic spin in diamond}, \href{http://dx.doi.org/10.1038/nature07279}{ {\it Nature (London)} \textbf{455}, 644 (2008)}.

\bibitem{Bal_08_NL} G. Balasubramanian, I. Y. Chan, R. Kolesov, M. Al-Hmoud, J. Tisler, C. Shin, C. Kim, A. Wojcik, P. R. Hemmer, A. Krueger, T. Hanke, A. Leitenstorfer, R. Bratschitsch, F. Jelezko, J. Wrachtrup, {\it Nanoscale imaging magnetometry with diamond spins under ambient conditions}, \href{http://dx.doi.org/10.1038/nature07278}{{\it Nature (London)} \textbf{455}, 648 (2008)}.

\bibitem{Tay_08_NP} J. M. Taylor, P. Cappellaro, L. Childress, L. Jiang, D. Budker, P. R. Hemmer, A. Yacoby, R. Walsworth, M. D. Lukin, {\it High-sensitivity diamond magnetometer with nanoscale resolution}, \href{http://dx.doi.org/10.1038/nphys1075}{{\it Nature Phys.} \textbf{4}, 810 (2008)}.

\bibitem{Dol_11_NP} F. Dolde, H. Fedder, M. W. Doherty, T. Nobauer, F. Rempp, G. Balasubramanian, T. Wolf, F. Reinhard, L. C. L. Hollenberg, F. Jelezko, J. Wrachtrup, {\it Electric-field sensing using single diamond spins}, \href{http://dx.doi.org/10.1038/nphys1969}{{\it Nature Phys.} \textbf{7}, 459 (2011)}.

\bibitem{Dol_14_PRL} F. Dolde, M. W. Doherty, J. Michl, I. Jakobi, B. Naydenov, S. Pezzagna, J. Meijer, P. Neumann, F. Jelezko, N. B. Manson, J. Wrachtrup, {\it Nanoscale Detection of a Single Fundamental Charge in Ambient Conditions Using the NV$^{-}$ Center in Diamond}, \href{http://link.aps.org/doi/10.1103/PhysRevLett.112.097603}{{\it Phys. Rev. Lett.} \textbf{112}, 097603 (2014)}.

\bibitem{Acosta_2010_PRL} V. M. Acosta, E. Bauch, M. P. Ledbetter, A. Waxman, L.-S. Bouchard, D. Budker, {\it Temperature Dependence of the Nitrogen-Vacancy Magnetic Resonance in Diamond}, \href{https://doi.org/10.1103/PhysRevLett.104.070801}{{\it Phys. Rev. Lett.} {\bf 104}, 070801 (2010)}.

\bibitem{Kucsko_Nature_2013} G. Kucsko, P. C. Maurer, N. Y. Yao, M. Kubo, H. J. Noh, P. K. Lo, H. Park, M. D. Lukin, {\it Nanometre-scale thermometry in a living cell}, \href{https://www.nature.com/nature/journal/v500/n7460/full/nature12373.html}{{\it Nature.} \textbf{500}, 54 (2013).}

\bibitem{Toy_13_PNASU} D. M. Toyli, C. F. de las Casas, D. J. Christle, V. V. Dobrovitski, D. D. Awschalom, {\it Fluorescence thermometry enhanced by the quantum coherence of single spins in diamond}, \href{http://www.pnas.org/content/110/21/8417.abstract}{{\it Proc. Natl. Acad. Sci. U.S.A.} \textbf{110}, 8417 (2013)}.

\bibitem{Neu_13_NL} P. Neumann, I. Jakobi, F. Dolde, C. Burk, R. Reuter, G. Waldherr, J. Honert, T. Wolf, A. Brunner, J. H. Shim, D. Suter, H. Sumiya, J. Isoya, J. Wrachtrup, {\it High-Precision Nanoscale Temperature Sensing Using Single Defects in Diamond}, \href{http://dx.doi.org/10.1021/nl401216y}{{\it Nano Lett.} \textbf{13}, 2738 (2013)}.

\bibitem{Doherty_14_PRL} M. W. Doherty, V. V. Struzhkin, D. A. Simpson, L. P. McGuinness, Y.-F. Meng, A. Stacey, T. J. Karle, R. J. Hemley, N. B. Manson, L. C. L. Hollenberg, S. Prawer, {\it Electronic Properties and Metrology Applications of the Diamond NV$^{-}$ Center under Pressure}, \href{https://doi.org/10.1103/PhysRevLett.112.047601} {{\it Phys. Rev. Lett.} {\bf 112}, 047601 (2014)}.

\bibitem{Cai NC14} J.-M. Cai, F. Jelezko, M. B. Plenio, {\it Hybrid sensors based on colour centres in diamond and piezoactive layers}, \href{https://www.nature.com/articles/ncomms5065}{{\it Nat. Commun.} \textbf{5}, 4065 (2014)}.

\bibitem{Schir14} R. Schirhagl, K. Chang, M. Loretz, C. L. Degen, {\it Nitrogen-Vacancy Centers in Diamond: Nanoscale Sensors for Physics and Biology}, \href{http://www.annualreviews.org/doi/abs/10.1146/annurev-physchem-040513-103659}{{\it Annu. Rev. Phys. Chem.} \textbf{65}, 83 (2014).}

\bibitem{Degen17} C. L. Degen, F. Reinhard, P. Cappellaro, {\it Quantum sensing}, \href{https://journals.aps.org/rmp/abstract/10.1103/RevModPhys.89.035002}{{\it Rev. Mod. Phys.} \textbf{89}, 035002 (2017).}

\bibitem{Hirose} M. Hirose, C. D. Aiello, P. Cappellaro, {\it Continuous dynamical decoupling magnetometry}, \href{https://doi.org/10.1103/PhysRevA.86.062320}{{\it Phys. Rev. A} {\bf 86}, 062320 (2012)}.

\bibitem{Fang2013} K. Fang, V. M. Acosta, C. Santori, Z. Huang, K. M. Itoh, H. Watanabe, S. Shikata, R. G. Beausoleil, {\it High-Sensitivity Magnetometry Based on Quantum Beats in Diamond Nitrogen-Vacancy Centers}, \href{https://doi.org/10.1103/PhysRevLett.110.130802}{{\it Phys. Rev. Lett.} {\bf 110}, 130802 (2013)}.

\bibitem{Cooper2014} A. Cooper, E. Magesan, H. Yum, P. Cappellaro, {\it Time-resolved magnetic sensing with electronic spins in diamond}, \href{https://www.nature.com/articles/ncomms4141}{{\it Nat. Commun.} {\bf 5}, 3141 (2014).}

\bibitem{Hall16} L. T. Hall, P. Kehayias, D. A. Simpson, A. Jarmola, A. Stacey, D. Budker, L. C. L. Hollenberg, {\it Detection of nanoscale electron spin resonance spectra demonstrated using nitrogen-vacancy centre probes in diamond}, \href{https://www.nature.com/articles/ncomms10211}{{\it Nat. Commun.} {\bf 7}, 10211 (2016)}.

\bibitem{Joas2017} T. Joas, A. M. Waeber, G. Braunbeck, F. Reinhard, {\it Quantum sensing of weak radio-frequency signals by pulsed Mollow absorption spectroscopy}, \href{https://doi.org/10.1038/s41467-017-01158-3}{{\it Nat. Commun.} {\bf 8}, 964 (2017)}.

\bibitem{Stark2017} A. Stark, N. Aharon, T. Unden, D. Louzon, A. Huck, A. Retzker, U. L. Andersen, F. Jelezko, {\it Narrow-bandwidth sensing of high-frequency fields with continuous dynamical decoupling}, \href{https://www.nature.com/articles/s41467-017-01159-2}{{\it Nat. Commun.} {\bf 8}, 1105 (2017)}.

\bibitem{Schmitt2017} S. Schmitt, T. Gefen, F. M. St\"urner, T. Unden, G. Wolff, C. M\"uller, J. Scheuer, B. Naydenov, M. Markham, S. Pezzagna, J. Meijer, I. Schwarz, M. Plenio, A. Retzker, L. P. McGuinness, F. Jelezko, {\it Submillihertz magnetic spectroscopy performed with a nanoscale quantum sensor}, \href{http://science.sciencemag.org/content/356/6340/832}{{\it Science} {\bf 356}, 832 (2017).}

\bibitem{Boss2017} J. M. Boss, K. S. Cujia, J. Zopes, C. L. Degen, {\it Quantum sensing with arbitrary frequency resolution}, \href{http://science.sciencemag.org/content/356/6340/837}{{\it Science} {\bf 356}, 837 (2017).}

\bibitem{Liu2017} Haibin Liu, Martin B. Plenio, and Jianming Cai, {\it Scheme for Detection of Single-Molecule Radical Pair Reaction Using Spin in Diamond}, \href{https://journals.aps.org/prl/abstract/10.1103/PhysRevLett.118.200402} {{\it Phys. Rev. Lett.} {\bf 118}, 200402 (2017)}.

\bibitem{Hsieh2019} S. Hsieh, P. Bhattacharyya, C. Zu, T. Mittiga, T. J. Smart, F. Machado, B. Kobrin, T. O. H\"{o}hn, N. Z. Rui, M. Kamrani, S. Chatterjee, S. Choi, M. Zaletel, V. V. Struzhkin, J. E. Moore, V. I. Levitas, R. Jeanloz, N. Y. Yao, {\it Imaging stress and magnetism at high pressures using a nanoscale quantum sensor}, \href{https://science.sciencemag.org/content/366/6471/1349}{{\it Science} {\bf 366}, 1349 (2019).}

\bibitem{Yip2019} King Yau Yip, Kin On Ho, King Yiu Yu, Yang Chen,  Wei Zhang, S. Kasahara, Y. Mizukami, T. Shibauchi, Y. Matsuda, Swee K. Goh, Sen Yang, {\it Measuring magnetic field texture in correlated electron systems under extreme conditions}, \href{https://science.sciencemag.org/content/366/6471/1355/tab-article-info}{{\it Science} {\bf 366}, 1355 (2019)}.

\bibitem{Lesik2019} Margarita Lesik, Thomas Plisson, L. Toraille, J. Renaud, F. Occelli, M. Schmidt, O. Salord, A. Delobbe, T. Debuisschert, L. Rondin, P. Loubeyre, J.-F. Roch, {\it Magnetic measurements on micrometer-sized samples under high pressure using designed NV centers}, \href{https://science.sciencemag.org/content/366/6471/1359}{{\it Science} {\bf 366}, 1359 (2019)}.

\bibitem{Bala09_NM} G. Balasubramanian, P. Neumann, D. Twitchen, M. Markham, R. Kolesov, N. Mizuochi, J. Isoya, J. Achard, J. Beck, J. Tissler, V. Jacques, P. R. Hemmer, F. Jelezko, J. Wrachtrup, {\it Ultralong spin coherence time in isotopically engineered diamond}, \href{https://www.nature.com/articles/nmat2420} {{\it Nature Materials.} \textbf{8}, 382-387 (2009).}

\bibitem{Lange_Sci_2010} G. de Lange, Z. H. Wang, D. Rist\`e, V. V. Dobrovitski, R. Hanson, {\it Universal Dynamical Decoupling of a Single Solid-State Spin from a Spin Bath},
\href{http://science.sciencemag.org/content/330/6000/60}{{\it Science} {\bf 330}, 60-63 (2010). }

\bibitem{Nay_PRB_2011} B. Naydenov, F. Dolde, L. T. Hall, C. Shin, H. Fedder, L. C. L. Hollenberg, F. Jelezko, J. Wrachtrup, {\it Dynamical decoupling of a single-electron spin at room temperature},
\href{https://journals.aps.org/prb/abstract/10.1103/PhysRevB.83.081201}{{\it Phys. Rev. B} {\bf 83}, 081201 (2011). }

\bibitem{Knowl14} H. S. Knowles, D. M. Kara, M. Ata\"ure. {\it Observing bulk diamond spin coherence in high-purity nanodiamonds}, \href{http://www.nature.com/nmat/journal/v13/n1/full/nmat3805.html}{{\it Nature Materials.} {\bf 13}, 21 (2014)}.

\bibitem{Mcgui11} L. P. McGuinness, Y. Yan, A. Stacey, D. A. Simpson, L. T. Hall, D. Maclaurin, S. Prawer, P. Mulvaney, J. Wrachtrup, F. Caruso, R. E. Scholten, L. C. Hollenberg, {\it Quantum measurement and orientation tracking of fluorescent nanodiamonds inside living cells}, \href{http:// DOI: 10.1038/NNANO.2011.64}{{\it Nature Nanotechnology.} {\bf 6}, 358 (2011)}.

\bibitem{Sage_Nature_2013} D. Le Sage	, K. Arai, D. R. Glenn, S. J. DeVience, L. M. Pham, L. Rahn-Lee, M. D. Lukin, A. Yacoby, A. Komeili, R. L. Walsworth, {\it Optical magnetic imaging of living cells}, \href{http://www.nature.com/nature/journal/v496/n7446/full/nature12072.html} {{\it Nature.} {\bf 496}, 486-489 (2013).}

\bibitem{Bala14} G. Balasubramanian, A. Lazariev, S. R. Arumugam, et al. {\it Nitrogen-Vacancy color center in diamond-emerging nanoscale applications in bioimaging and biosensing}, \href{https://www.ncbi.nlm.nih.gov/pubmed/24875635}{{\it Current Opinion in Chemical Biology.} {\bf 20}, 69-77 (2014)}.

\bibitem{Wu2016} Y. Wu, F. Jelezko, M.B. Plenio, T. Weil, {\it Diamond Quantum Devices in Biology}, \href{http://onlinelibrary.wiley.com/doi/10.1002/anie.201506556/full}{{\it Angewandte Chemie - International Edition} {\bf 55}, 6586 - 6598 (2016)}.

\bibitem{Kurizki} G. Gordon, G. Kurizki, D. A. Lidar. {\it Optimal Dynamical Decoherence Control of a Qubit}, Phys. Rev. Lett. {\bf 101}, 010403 (2008)

\bibitem{Cai_NJP_2012} J.-M. Cai, B. Naydenov, R. Pfeiffer, L. McGuinness, K. Jahnke, F. Jelezko, M.B. Plenio, A. Retzker, {\it Robust dynamical decoupling with concatenated continuous driving}, \href{http://iopscience.iop.org/article/10.1088/1367-2630/14/11/113023/meta;jsessionid=66B1A2EA818E2A18AEBC591BF41189EF.c1.iopscience.cld.iop.org}{{\it New J. Phys.} {\bf 14}, 113023 (2012)}.

\bibitem{Cohen16} I. Cohen, N. Aharon, A. Retzker, {\it Continuous dynamical decoupling utilizing time-dependent detuning}, \href{http://onlinelibrary.wiley.com/doi/10.1002/prop.201600071/full}{{\it Fortschr. Phys.} {\bf 64}, 1521 (2016)}.

\bibitem{Farf17} D. Farfurnik, N. Aharon, I. Cohen, Y. Hovav, A. Retzker, N. Bar-Gill, {\it Experimental realization of time-dependent phase-modulated continuous dynamical decoupling}, \href{https://journals.aps.org/pra/abstract/10.1103/PhysRevA.96.013850}{{\it Phys. Rev. A} {\bf 96}, 013850 (2017)}.

\bibitem{Shu2018} Zijun Shu, Yu Liu, Qingyun Cao, Pengcheng Yang, Shaoliang Zhang, Martin B. Plenio, Fedor Jelezko, and Jianming Cai, {\it Observation of Floquet Raman Transition in a Driven Solid-State Spin System}, \href{https://doi.org/10.1103/PhysRevLett.121.210501}{{\it Phys. Rev. Lett.}{\bf 121}, 210501 (2018)}.

\bibitem{Yu2019} M. Yu, P.-C. Yang, M.-S. Gong, Q.-Y. Cao, Q.-Y. Lu, H.-B. Liu, M. B. Plenio, F. Jelezko, T. Ozawa, N. Gold- man, S.-L. Zhang, and J.-M. Cai, {\it Experimental measurement of the quantum geometric tensor using coupled qubits in diamond}, \href{https://doi.org/10.1093/nsr/nwz193}{{\it Natl. Sci. Rev.} nwz193, 10.1093/nsr/nwz193 (2019)}.

\bibitem{Jac16} P. Jamonneau, M. Lesik, J. P. Tetienne, I. Alvizu, L. Mayer, A. Dr\'eau, S. Kosen, J.-F. Roch, S. Pezzagna, J. Meijer, T. Teraji, Y. Kubo, P. Bertet, J. R. Maze, V. Jacques, {\it Competition between electric field and magnetic field noise in the decoherence of a single spin in diamond}, \href{https://journals.aps.org/prb/abstract/10.1103/PhysRevB.93.024305}{{\it Phys. Rev. B} {\bf 93}, 024305 (2016)}.

\bibitem{Medford_PRL_2012} J. Medford, \L{}. Cywi\'nski, C. Barthel, C. M. Marcus, M. P. Hanson, A. C. Gossard, {\it Scaling of dynamical decoupling for spin qubits}
\href{https://journals.aps.org/prl/abstract/10.1103/PhysRevLett.108.086802}{{\it Phys. Rev. Lett.} {\bf 108}, 086802 (2012).}

\bibitem{Ahmed_2013_PRA} M. A. Ali Ahmed, G. A. \'Alvarez, D. Suter, {\it Robustness of dynamical decoupling sequences}, \href{https://doi.org/10.1103/PhysRevA.87.042309}{{\it Phys. Rev. A} {\bf 87}, 042309 (2013)}.

\bibitem{Nati}
N. Aharon, I. Cohen, F. Jelezko, A. Retzker. {\it Fully robust qubit in atomic and molecular three-level systems},
\href{http://iopscience.iop.org/article/10.1088/1367-2630/aa4fd3/meta}{{\it New J. Phys.} \textbf{18}, 123012 (2016)}.

\bibitem{nir}
N. Bar-Gill, L.M. Pham, A. Jarmola, D. Budker, R.L. Walsworth, {\it Solid-state electronic spin coherence time approaching one second}, \href{https://www.nature.com/articles/ncomms2771}{{\it Nat. Commun.} {\bf 4}, 1743 (2013).}

\bibitem{Woo2000} Im-Sun Woo, In-Koo Rhee, and Heui-Dong Park, {\it Differential Damage in Bacterial Cells by Microwave Radiation on the Basis of Cell Wall Structure}, \href{https://doi.org/10.1128/AEM.66.5.2243-2247.2000}{{\it Appl. Environ. Microbiol.} {\bf 66}, 2243 (2000).}

\bibitem{Shamis2012} Y. Shamis, R. Croft, A. Taube, R. J. Crawford, E. P. Ivanova, {\it Review of the specific effects of microwave radiation on bacterial cells}, \href{https://doi.org/10.1007/s00253-012-4339-y} {{\it Appl Microbiol Biotechnol} {\bf 96}, 319 (2012).}

\bibitem{Loretz15} M. Loretz, J. M. Boss, T. Rosskopf, H. J. Mamin, D. Rugar, C. L. Degen, {\it Spurious Harmonic Response of Multipulse Quantum Sensing Sequences}, \href{https://journals.aps.org/prx/abstract/10.1103/PhysRevX.5.021009}{{\it Phys. Rev. X} {\bf 5}, 021009 (2015).}

\bibitem{Haase16}J. F. Haase, Z.-Y. Wang, J. Casanova, M. B. Plenio, {\it Pulse-phase control for spectral disambiguation in quantum sensing protocols}, \href{https://journals.aps.org/pra/abstract/10.1103/PhysRevA.94.032322}{{\it Phys. Rev. A} {\bf 94}, 032322 (2016).}

\bibitem{Shu2017} Z.-J. Shu, Z.-D. Zhang, Q.-Y. Cao, P.-C. Yang, M. B. Plenio, C. M\"uller, J. Lang, N. Tomek, B. Naydenov, L. P. McGuinness, F. Jelezko, J.-M. Cai, {\it Unambiguous nuclear spin detection using an engineered quantum sensing sequence}, \href{https://journals.aps.org/pra/abstract/10.1103/PhysRevA.96.051402}{{\it Phys. Rev. A} {\bf 96}, 051402 (2017).}

\end{thebibliography}

\end{document}